\documentclass{article}
\usepackage{graphicx}
\pdfoutput=1
\usepackage{epsfig}
\usepackage{amsfonts}
\usepackage{amssymb}
\usepackage{amsmath}
\usepackage{epsf}
\usepackage{multirow}
\usepackage{array}
\usepackage{cite}
\usepackage{authblk}
\allowdisplaybreaks
 \usepackage{amssymb}
\usepackage{enumitem}  
\usepackage{amsmath}
\usepackage{subfig}
\usepackage{hyperref}
\usepackage{cleveref}
\usepackage{float}
\usepackage{color}

\date{}

\newcommand{\insertplot}[5]{\begin{figure}
 \hfill\hbox to 0.05in{\vbox to #5in{\vfill
 \inputplot{#1}{#4}{#5}}\hfill}
 \hfill\vspace{-.1in}
 \caption{#2}\label{#3}
 \end{figure}}
 \newcommand{\inputplot}[3]{
 \special{ps: plotfile #1}
}
\newcounter{fig}

\newcommand{\ee}{\end{equation}}
\newcommand{\eea}{\end{eqnarray}}
\newcommand{\be}{\begin{equation}}
\newcommand{\bea}{\begin{eqnarray}}

\numberwithin{equation}{section}

\begin{document}

\title{ \textbf{
Bifurcations in Bosonic Stars:\\
 chains and rings from spherical solutions
}
} 

\author[1]{Chen Liang\footnote{liang.c@ua.pt}}
\author[1]{Carlos A. R. Herdeiro\footnote{herdeiro@ua.pt}}
\author[1]{Eugen Radu\footnote{eugen.radu@ua.pt}}

\affil[1]{\normalsize Departamento de Matemática da Universidade de Aveiro and 

Center for Research and Development in Mathematics and Applications -- CIDMA

Campus de Santiago, 3810-19 3 Aveiro, Portugal}

\maketitle

\begin{abstract}
  We study the bifurcation phenomena between spherical and axisymmetric bosonic stars. By numerically solving for the zero-modes of spherical bosonic stars under specific axially symmetric perturbations, we discover that excited state spherical bosonic stars bifurcate into two types of axisymmetric bosonic stars under $\ell=2$ perturbations, with matter distributions resembling chains and rings, respectively. Meanwhile, $\ell=4$ axisymmetric perturbations lead spherical scalar bosonic stars to bifurcate into a new type of axisymmetric bosonic stars, exhibiting a mixed chain-like and ring-like matter distribution, which we refer to as gyroscope-like. Additionally, for the first time, we have constructed chains of scalar bosonic stars with 7 constituents and their corresponding ring-like scalar bosonic stars. Our results provide an explanation for the bifurcations in bosonic stars from the perspective of perturbations, and by analyzing physical quantities such as quadrupoles and energy densities we systematically discuss the impact of axisymmetric perturbations on spherical bosonic stars.
  \end{abstract}

 %%%%%%%%%%%%%%%%%%%%%%%%%%%%%%%%%%%%%%%%%%%
 \section{Introduction}
 %%%%%%%%%%%%%%%%%%%%%%%%%%%%%%%%%%%%%%%%%%%
An equilibrium physical system may lose its stability under specific perturbations, leading to changes in its equilibrium characteristics. Consequently, stationary solutions to the equations describing this system may bifurcate from the main branch, creating new branches. This phenomenon of bifurcation is widespread in areas of physics such as nonlinear dynamics \cite{Chin:1994cd,Brand:1989cd,Crawford:1991cd}, statistical physics \cite{Chechkin:2003cd}, fluid dynamics \cite{Basaran:1989cd,Duft:2003cd,Burton:2011cd}, laser physics \cite{Giudici:1997cd}, and quantum mechanics \cite{Schleich:1989cd,Hansen:1989cd}.

Bifurcation also occurs in theories of gravity. For example, in higher-dimensional General Relativity, black strings in spacetime dimensions $D>4$ are unstable against long-wavelength perturbations, known as the Gregory-Laflamme instability \cite{Gregory:1993vy,Kleihaus:2013yaa,Wiseman:2002zc,Sorkin:2004qq}. This leads to a bifurcation into a family of non-uniform black string solutions at the onset of the instability. In certain modified theories of gravity, compact celestial bodies such as neutron stars \cite{Damour:1992we,Damour:1993hw} and black holes \cite{Doneva:2017bvd,Silva:2017uqg,Herdeiro:2018wub} undergo spontaneous scalarization under strong spacetime curvature, leading to new branches that bifurcate from the original model's trunk family. Moreover, in studies concerning hairy black holes, the hairy generalizations of vacuum black holes bifurcate from the conventional black holes at the threshold of superradiance \cite{Hod:2012px,Herdeiro:2014goa,Wang:2018xhw,Degollado:2018ypf}.

Recently, bifurcations have been observed in another type of compact object, \textit{ bosonic stars} (BSs) \cite{Herdeiro:2021mol,Sun:2022duv,Herdeiro:2023wqf}. BSs are solitonic solutions described by massive bosonic fields with integer spins, coupled to gravity. In 1968, Kaup first constructed numerical solutions for a system of massive complex scalar fields coupled to Einstein's gravity, known as \textit{boson stars} \cite{Kaup:1968zz}. Vector BSs, or \textit{Proca stars}, were proposed almost half a century later by Brito \textit{et al.} \cite{Brito:2015pxa}. Heisenberg's uncertainty principle prevents BSs from collapsing under their own gravitational attraction. BSs are considered candidates for dark matter \cite{Matos:1999et,Matos:2000ss,Hu:2000ke,Hui:2016ltb} and can also mimic black holes \cite{Cardoso:2019rvt,Glampedakis:2017cgd,Herdeiro:2021lwl}. In recent developments, a novel axially symmetric configuration of BSs was proposed, where multiple BSs were symmetrically aligned along the symmetry axis around the origin, called \textit{ chain of BSs} \cite{Herdeiro:2020kvf,Herdeiro:2021mol,Sun:2022duv,Gervalle:2022fze,Herdeiro:2023wqf}. Similar chain-like configurations are also present in flat space-time analogs of BSs, known as \textit{Q-balls} \cite{Loiko:2020htk}, and have been extensively studied in other soliton models such as Skyrmions \cite{Krusch:2004uf,Shnir:2009ct,Shnir:2015aba,Shnir:2021gki}, sphalerons \cite{Kleihaus:2008gn,Ibadov:2008hj,Ibadov:2010ei,Ibadov:2010hm,Teh:2014saa} and dyons \cite{Kleihaus:1999sx,Kleihaus:2000hx,Kleihaus:2003nj,Kleihaus:2003xz,Kleihaus:2004is,Kleihaus:2004fh,Teh:2004bq,Paturyan:2004ps,Kleihaus:2005fs,Kunz:2006ex,Kunz:2007jw,Lim:2011ra,Teh:2014zea}. For static chains of BSs, besides their unique symmetry compared to typical spherical BSs (SBSs), an interesting feature is that their solution space bifurcates from the one of SBSs, breaking the system's spherical symmetry at the bifurcation point. This bifurcation phenomenon merits further discussion. 

Furthermore, the astrophysical viability of BSs as theoretical celestial models critically depends on their stability. The linear \cite{Gleiser:1988rq,Jetzer:1988vr,Lee:1988av,Gleiser:1988ih,Kojima:1991np,Yoshida:1994xi,Macedo:2013jja,Santos:2024vdm} and nonlinear \cite{Seidel:1990jh,Balakrishna:1997ej,Guzman:2004jw,Kain:2021rmk,Sanchis-Gual:2021phr,Siemonsen:2020hcg,Sanchis-Gual:2017bhw,Herdeiro:2023wqf} stabilities of BSs have been extensively studied. Yet, research on linear perturbations of BSs has primarily focused on spherically symmetric perturbations, with studies related to axially symmetric perturbations being relatively scarce.

The objective of this paper is to analyze the bifurcations between static axisymmetric BSs (ABSs) and SBSs using linear axially symmetric perturbation methods. We find that excited state SBSs exhibit zero modes under specific axially symmetric perturbations, at which point SBSs bifurcate to form new ABSs. Unlike typical bifurcations, those in BSs point towards two distinct branches, associated with the chain-like and ring-like configurations of BSs mentioned in \cite{Sun:2022duv}. We analyze the bifurcations of BSs under two types of axially symmetric perturbation modes and construct some new models of ABSs.

This paper is organized as follows. In Section \ref{sec:models}, we introduce two models of bosonic fields with different spins minimally coupled to Einstein's gravity, the ansatz and the ordinary differential equations (ODEs) and partial differential equations (PDEs) used to construct BSs. Section \ref{sec:bc} describes the boundary conditions satisfied by the models and some physical quantities. We also detail scaling and numerical methods. Section \ref{sec:bif} briefly presents some bifurcations in BSs and provides the ansatz for the considered axisymmetric perturbations along with the corresponding perturbation equations. Numerical results are presented in Section \ref{sec:result}, explaining the bifurcations in BSs from a perturbative perspective. The paper concludes with a summary of findings and poses several open questions in Section \ref{conclusions}.

 %%%%%%%%%%%%%%%%%%%%%%%%%%%%%%%%%%%%%%%%%%%
 
 \section{The framework}
 \label{sec:models}
\subsection{The models and ansatz}

 %%%%%%%%%%%%%%%%%%%%%%%%%%%%%%%%%%%%%%%%%%%
  
We consider the action for Einstein's gravity minimally coupled to a complex massive spin-$s$ field, with $s=0,1$:
\begin{equation}        
\label{action}
 S=  \int  d^4x\sqrt{-g}\left( \frac{R}{16 \pi G} + \mathcal{L}_s
 \right)\,,
\end{equation}
where the matter Lagrangians are as follows:
\begin{equation}        
  \label{lag}
  \mathcal{L}_0= -g^{ab}\Psi_{,a}\bar{\Psi}_{,b} - \mu^2|\Psi|^2\,,\qquad 
  \mathcal{L}_1= -\frac{1}{4}\mathcal{F}_{ab}\bar{\mathcal{F}}^{ab}-\frac{1}{2}\mu^2\mathcal{A}_a\bar{\mathcal{A}}^a\,,
\end{equation}
where $\Psi$ is a complex scalar field, $\mathcal{A}$ is a complex Proca field and $\mathcal{F} = d\mathcal{A}$; $\mu$ is the mass of each bosonic field. 
The resulting field equations are:
\begin{subequations}\label{allequ}
\begin{align}
 E_{ab}^{(s)}=
R_{ab}-\frac{1}{2}g_{ab}R - 8\pi G T^{(s)}_{ab} & = 0\,, \label{E-eq}\\
\nabla^2 \Psi -\mu^2\Psi & = 0\,, \label{KG-eq}\\
\nabla_a\mathcal{F}^{ab} - \mu^2 \mathcal{A}^b & = 0 
\label{Proca-eq}.
\end{align}  
\end{subequations}
Also, the energy-momentum tensors read
\begin{subequations}
\begin{align}
T^{(0)}_{ab} & =  
\bar{\Psi}_{,a}\Psi_{,b} + \bar{\Psi}_{,b} \Psi_{,a}
%+\Psi_{,b}^*\Psi_{,a} 
-g_{ab} 
\left[ 
  \frac{1}{2} g^{cd}
  \left(
    \bar{\Psi}_{,c}\Psi_{,d} + \bar{\Psi}_{,d} \Psi_{,c}
  \right)
 %+\Psi_{,d}^*\Psi_{,c} )
+ \mu^2 |\Psi|^2
\right]\,,\\
T^{(1)}_{ab} & = \frac{1}{2} g^{cd} \left( \mathcal{F}_{ac} \bar{\mathcal{F}}_{bd} + \bar{\mathcal{F}}_{ac}\mathcal{F}_{bd} \right) - \frac{1}{4}g_{ab}\mathcal{F}_{cd} \bar{\mathcal{F}}^{cd} + \frac{1}{2} \mu^2 \left( \mathcal{A}_a \bar{\mathcal{A}}_b + \bar{\mathcal{A}}_a \mathcal{A}_b - g_{ab}\mathcal{A}_c \bar{\mathcal{A}}^c \right)\,.
\end{align}
\end{subequations}
 In this work, we consider a general axially symmetric line element:
\begin{equation}
  ds^2 =g_{tt}(r,\theta) dt^2 + 
  g_{rr}(r,\theta)dr^2 + g_{\theta\theta}(r,\theta)d\theta^2+ g_{\varphi\varphi}(r,\theta) d\varphi^2\,.
\end{equation}

A substantial body of literature already exists on the SBSs. To construct the SBSs, a spherically symmetric metric of the following form is employed:
\begin{equation}\label{ansatz:smetric}
  g_{tt} =- N(r)\sigma^2(r)\,, \quad 
  g_{rr} = \frac{1}{N(r)}\,, \quad 
  % g_{tt} =- \left(1-\frac{2m(r)}{r} \right)\sigma^2(r)\,, \quad 
  % g_{rr} = \frac{1}{1-\frac{2m(r)}{r}}\,, \quad 
  g_{\theta\theta} = r^2\,, \quad 
  g_{\varphi\varphi} = r^2\sin^2 \theta\,, \quad 
  \mbox{with}\quad N(r)=1-\frac{2m(r)}{r}\,,
\end{equation}
where the metric functions $m(r)$ and $\sigma(r)$ depend only on $r$. This form of the metric simplifies the derivation of the perturbation equations that we need to solve later.

As for the ABSs, discussed in \cite{Herdeiro:2020kvf} and \cite{Herdeiro:2023wqf}, we adopt the following metric:
\begin{equation}\label{ansatz:ametric}
  g_{tt} =- e^{2F_0(r,\theta)}\,,\quad
  g_{rr} = e^{2F_1(r,\theta)}\,,\quad
  g_{\theta\theta} = r^2 e^{2F_1(r,\theta)}\,,\quad
  g_{\varphi\varphi} = r^2 \sin^2\theta e^{2F_2(r,\theta)} \,.
\end{equation}
The matter fields $\Psi $ and $\mathcal{A}$
ansatz
have the following form
\begin{subequations}\label{ansatz:bg}
\begin{align}
\Psi & =\psi(r,\theta)e^{ -i\omega t }\,, \\
\mathcal{A}_a dx^a & = \left[ V(r,\theta)dt + iH_1(r,\theta)dr + iH_2(r,\theta)d\theta \right] e^{ -i\omega t }\,,
\end{align}
\end{subequations}
where $\psi$, $V$, $H_1$ and $H_2$ are real functions. In the spherical case, these matter functions are independent of angular $\theta$, with $H_2 = 0$.
 
%%%%%%%%%%%%%%%%%%%%%%%%%%%%%%%%%%%%%%%%%%%
\subsection{The equations of motion}\label{teom}
%%%%%%%%%%%%%%%%%%%%%%%%%%%%%%%%%%
Substituting (\ref{ansatz:smetric}) and (\ref{ansatz:bg}) into (\ref{allequ}) yields the following ODEs for constructing the SBSs:
\begin{subequations}\label{bgequ_spherical}
  \begin{align}
    & \psi''+ 
    \left(\frac{N'}{N}+\frac{\sigma'}{\sigma}+\frac{2}{r}\right)\psi'+
    \left(\frac{\omega^2}{N^2 \sigma^2}-\frac{\mu^2}{N}\right)\psi=0\,,
    \label{bgequ_spherical_1}\\
    & m'-
    4\pi G r^2\left[  N \psi'^2+
     \left(\mu^2+\frac{\omega^2}{N \sigma^2}\right)\psi^2 \right]=0\,,
     \label{bgequ_spherical_2}\\
    & \sigma'- 
    8\pi G r\left[  \sigma \psi'^2+
    \frac{ \omega^2 \psi^2}{N^2 \sigma} \right] =0\,,
    \label{bgequ_spherical_3}\\
    & \frac{\partial}{\partial r}\left[ \frac{r^2}{\sigma} \left( V'-\omega H_1 \right) \right] -
    \frac{\mu^2 r^2 V}{\sigma N} = 0\,,
    \label{bgequ_spherical_4}\\
    & V' + \left( \frac{\mu^2 \sigma^2 N}{\omega} - \omega \right) H_1 = 0\,,
    \label{bgequ_spherical_5}\\
    & m'-
    2\pi G r^2\left[ \frac{\left( V'-\omega H_1 \right)^2}{\sigma^2} + \mu^2\left(  \frac{V^2}{N\sigma^2} + NH_1^2 \right) \right]=0\,,
    \label{bgequ_spherical_6}\\
    & \sigma'-
    4\pi Gr\mu^2 \left( \frac{V^2}{N^2\sigma^2} + H_1^2 \right)=0\,,
    \label{bgequ_spherical_7}
\end{align}
\end{subequations}
where the prime refers to derivative with respect to $r$. The ODEs (\ref{bgequ_spherical_1}--\ref{bgequ_spherical_3}) and (\ref{bgequ_spherical_4}--\ref{bgequ_spherical_7}) can be solved numerically to construct scalar and Proca SBSs, respectively.

For the axisymmetric models, the explicit forms of the equations are much more complicated. The equations (\ref{allequ}) and ansatz (\ref{ansatz:ametric}-\ref{ansatz:bg}) lead to the following PDEs:
\begin{subequations}\label{bgequ_axial}
  \begin{align}
& e^{2 F_1}\phi \left( \omega^2 e^{-2 F_0} - \mu ^2\right)+
\frac{\phi _{,\theta} \left(F_{0,\theta}+F_{2,\theta}+\cot\theta\right)}{r^2}+\phi_{,r} \left(F_{0,r}+F_{2,r}+\frac{2}{r}\right)+\frac{\phi _{,\theta \theta}}{r^2}+\phi _{,rr}=0
\,,\label{bgequ_axial_1}\\
\notag\\
& 8 \pi G e^{2 F_1}  \phi ^2 \left(\mu ^2-2 e^{-2 F_0} \omega^2\right)+\frac{F_{0,\theta}\left(F_{0,\theta}+F_{2,\theta}+\cot\theta\right)}{r^2}+F_{0,r}\left(F_{0,r}+F_{2,r}+\frac{2}{r}\right)+\frac{F_{0,\theta \theta}}{r^2}+F_{0,rr}=0
\,,\label{bgequ_axial_2}\\
& 8 \pi  G \left(\phi ^2 \omega^2 e^{2 F_1-2 F_0}+\frac{\phi _{,\theta}^2}{r^2}+\phi_{,r}^2\right)-\frac{F_{0,\theta} \left(F_{2,\theta}+\cot\theta\right)}{r^2}-F_{0,r} \left(F_{2,r}+\frac{1}{r}\right)+\frac{F_{1,\theta \theta}}{r^2}+\frac{F_{1,r}}{r}+F_{1,rr}=0
\,,\label{bgequ_axial_3}\\
\notag\\
& \frac{F_{2,\theta} \left(F_{0,\theta}+2 \cot\theta+ F_{2,\theta} \right)}{r^2}+F_{2,r}\left(F_{0,r}+F_{2,r}+\frac{3}{r}\right)+\frac{F_{0,\theta} \cot\theta}{r^2}+\frac{F_{0,r}}{r}+8 \pi G  e^{2 F_1} \mu ^2 \phi ^2+\frac{F_{2,\theta\theta }}{r^2}+F_{2,rr}=0
\,,\label{bgequ_axial_4}\\
\notag\\
& e^{2 F_1} V \left(e^{-2 F_0} \omega^2-\mu ^2\right)+\frac{V_{,\theta} \left(F_{2,\theta}-F_{0,\theta}+\cot\theta\right)}{r^2}+V_{,r}\left(F_{2,r}-F_{0,r}+\frac{2}{r}\right)
\notag\\
& \qquad\qquad\qquad\qquad\qquad\qquad\qquad\qquad\qquad\qquad\qquad
 -\frac{2 \omega \left(H_1 rF_{0,r}+H_2 F_{0,\theta}\right)}{r^2}+\frac{V_{,\theta \theta}}{r^2}+V_{,rr}=0
\,,\label{bgequ_axial_5}\\
\notag\\
& H_1 \left[e^{2F_1} \left(e^{-2 F_0} \omega^2-\mu^2\right)-2 F_{1,r} \left(F_{0,r}+F_{2,r}+\frac{1}{r}\right)-\frac{F_{0,r}}{r}+F_{0,rr}-\frac{F_{2,r}}{r}+F_{2,rr}-\frac{2}{r^2}\right]+
\notag\\
& \frac{H_{1,\theta} \left(F_{0,\theta}-2F_{1,\theta}+F_{2,\theta}+\cot\theta\right)}{r^2}+H_{1,r}\left(F_{0,r}-2 F_{1,r}+F_{2,r}\right)+
\frac{H_2 \left(r F_{0,r\theta}-2F_{0,\theta}-2 F_{2,\theta}+rF_{2,r\theta}-2 \cot\theta\right)}{r^2}
\notag\\
& +\frac{2 r F_{1,\theta} H_{2,r}-2H_{2,\theta}}{r^2}
-\frac{2F_{1,r}\left[H_2\left(F_{0,\theta}+F_{2,\theta}+\cot\theta\right)+H_{2,\theta}\right]}{r}+2 r V \omega e^{2F_1-2 F_0} F_{0,r}+\frac{H_{1,\theta \theta }}{r^2}+H_{1,rr}=0
\,,\label{bgequ_axial_6}\\
\notag\\
& 
\frac{H_2 \left[e^{2 F_1} r^2 \left(e^{-2 F_0} \omega^2-\mu ^2\right)-2F_{1,\theta} \left(F_{0,\theta}+F_{2,\theta}+\cot \theta\right)+F_{0,\theta \theta }+F_{2\theta\theta }-\csc ^2\theta\right]}{r^2}
\notag\\
& 
-\frac{2 F_{1,\theta} \left[H_1 \left(r F_{0,r}+rF_{2,r}+1\right)+r H_{1,r}\right]}{r^2}+\frac{H_{2,\theta} \left(F_{0,\theta}-2F_{1,\theta}+F_{2,\theta}+\cot\theta\right)}{r^2}+H_{2,r}\left(F_{0,r}-2 F_{1,r}+F_{2,r}\right)
\notag\\
& \qquad\qquad\qquad
+2 V \omega e^{2 F_1-2 F_0}F_{0,\theta}+\frac{H_1 \left(F_{0,r\theta}+F_{2\text{r$\theta$}}\right)}{r}+\frac{2 H_{1,\theta} \left(rF_{1,r}+1\right)}{r^2}+\frac{H_{2,\theta \theta }}{r^2}+H_{2,rr}=0
\,,\label{bgequ_axial_7}\\
\notag\\
& \frac{F_{0,\theta} \left(F_{0,\theta}+F_{2,\theta}+\cot\theta\right)}{r^2}
+F_{0,r}\left(F_{0,r}+F_{2,r}+\frac{2}{r}\right)+\frac{F_{0,\theta \theta}}{r^2}+F_{0,rr}
-4 \pi  G \Bigg\{
  2 \mu ^2 V^2 e^{2 F_1-2 F_0}
\notag\\
& +\frac{e^{-2 F_0} \left[\omega^2\left(H_1^2+H_2^2\right)+2 H_1 r \omega V_{,r}+2 H_2 \omega V_{,\theta}+r^2V_{,r}^2+V_{,\theta}^2\right]}{r^2}+\frac{e^{-2F_1}\left(H_{1,\theta}-rH_{2,r}\right){}^2}{r^4}
 \Bigg\}=0
 \,,\label{bgequ_axial_8}\\
 \notag\\
& 4 \pi  G \left[\mu ^2 V^2 e^{2 F_1-2 F_0}+\frac{2 e^{-2 F_1} \left(H_{1,\theta}-rH_{2,r}\right){}^2}{r^4}+\frac{\mu ^2\left(H_1^2+H_2^2\right)}{r^2}\right]-\frac{F_{0,\theta}\left(F_{2,\theta}+\cot\theta \right)}{r^2}
\notag\\
& \qquad\qquad\qquad\qquad\qquad\qquad\qquad\qquad\qquad\qquad\quad
-F_{0,r}\left(F_{2,r}+\frac{1}{r}\right)+\frac{F_{1,\theta \theta}}{r^2}+\frac{F_{1,r}}{r}+F_{1,rr}=0
\,,\label{bgequ_axial_9}\\
\notag\\
& 4 \pi  G \left\{e^{-2 F_0} \left[\frac{\omega^2 \left(H_1^2+H_2^2\right)}{r^2}+\frac{V_{,r} \left(2H_1 \omega+V_{,r}\right)}{r}+\frac{V_{,\theta} \left(2 H_2 \omega+V_{,\theta}\right)}{r^2}\right]-\frac{e^{-2 F_1} \left(H_{1,\theta}-rH_{2,r}\right){}^2}{r^4}\right\}+
\notag\\
& \quad
\frac{F_{2,\theta} \left(F_{0,\theta}+F_{2,\theta}+2\cot\theta\right)}{r^2}+F_{2,r}\left(F_{0,r}+F_{2,r}+\frac{3}{r}\right)+\frac{F_{0,\theta} \cot\theta+r F_{0,r}}{r^2}+\frac{F_{2,\theta \theta }}{r^2}+F_{2,rr}=0\,,\label{bgequ_axial_10}
  \end{align}
\end{subequations}
where (\ref{bgequ_axial_1}) and (\ref{bgequ_axial_5}-\ref{bgequ_axial_7}) are obtained from (\ref{KG-eq}) and (\ref{Proca-eq}), respectively. The equations (\ref{bgequ_axial_2}-\ref{bgequ_axial_4}) and (\ref{bgequ_axial_8}-\ref{bgequ_axial_10}), which represent the Einstein equations that we are numerically solving for the scalar and Proca cases, are obtained from the following combinations of the Einstein equations:
\begin{equation}
  \begin{aligned}
  & E_r^{r(s)} + E_\theta^{\theta(s)} + E_\varphi^{\varphi(s)} - E_t^{t(s)} = 0
  \,,\\
  & E_r^{r(s)} + E_\theta^{\theta(s)} - E_\varphi^{\varphi(s)} - E_t^{t(s)} = 0
  \,,\\
  & E_r^{r(s)} + E_\theta^{\theta(s)} - E_\varphi^{\varphi(s)} + E_t^{t(s)} = 0
 \,.
  \end{aligned} 
  \end{equation} 
%%%%%%%%%%%%%%%%%%%%%%%%%%%%%%%%%%%%%%%%%%
Additionally, there are two constraint equations for both the scalar and Proca cases to verify accuracy:
\begin{equation}
  \begin{aligned}
  E_r^{r(s)} - E_\theta^{\theta(s)} = 0
  \,,\qquad 
  E_r^{\theta(s)}  = 0
  \,.
\end{aligned} 
\end{equation} 
The explicit forms of the above constraint equations are:
\begin{subequations}\label{constraint}
  \begin{align}
  & F_{1,r} \left(2 F_{2,r}+\frac{2}{r}\right)+F_{0,r} \left(2F_{1,r}+\frac{1}{r}\right)-F_{1,\theta} \left(\frac{2 F_{2,\theta}}{r^2}+\frac{2 \cot\theta}{r^2}\right)-\frac{2 F_{1,\theta} F_{0,\theta}}{r^2}+\frac{2 F_{2,\theta} \cot\theta}{r^2}
  \notag\\
  & +\frac{F_{2,\theta}^2}{r^2}+\frac{F_{2,\theta\theta}+F_{0,\theta\theta}}{r^2}+\frac{F_{0,\theta}^2}{r^2}-\frac{F_{2,r}}{r}-F_{2,r}^2-F_{2,rr}-F_{0,r}^2-F_{0,rr}+16 \pi  G \left(\frac{\psi _{,\theta}^2}{r^2}- \psi _{,r}^2\right)=0
  \,,\label{constraint_1}\\
  & \notag\\
  & F_{0,\theta} \left(F_{1,r}+\frac{1}{r}\right)+F_{1,r} \left(F_{2,\theta}+\cot\theta\right)+F_{1,\theta}\left(F_{2,r}+\frac{1}{r}\right)+F_{0,r}\left(F_{1,\theta}-F_{0,\theta}\right)-F_{2,r} \left(F_{2,\theta}+\cot\theta\right)
  \notag\\
  & \qquad\qquad\qquad\qquad\qquad\qquad\qquad\qquad\qquad\qquad\qquad\qquad\qquad
   -F_{2,r\theta}-F_{0,r\theta}-16 \pi  G \psi _{,r} \psi _{,\theta}=0
   \,,\label{constraint_2}\\
   & \notag\\
  & F_{1,r} \left(2F_{2,r}+\frac{2}{r}\right)+F_{0,r} \left(2 F_{1,r}+\frac{1}{r}\right)-F_{1,\theta}\left(\frac{2 F_{2,\theta}}{r^2}+\frac{2 \cot\theta}{r^2}\right)-\frac{2 F_{1,\theta}F_{0,\theta}}{r^2}+\frac{2 F_{2,\theta} \cot \theta}{r^2}
  \notag\\
  &+\frac{F_{2,\theta}^2}{r^2}+\frac{F_{2,\theta\theta}+F_{0,\theta\theta}}{r^2}+\frac{F_{0,\theta}^2}{r^2}-\frac{F_{2,r}}{r}-F_{2,r}^2-F_{2,rr}-F_{0,r}^2-F_{0,rr}
  \notag\\
  & -\frac{8 \pi  G e^{-2 F_0} \left[H_1^2 \left(\mu ^2 e^{2 F_0}-\omega^2\right)+H_2^2 \left(\omega^2-\mu ^2 e^{2 F_0}\right)-2 r \omega H_1 V_{,r}+2 \omega H_2V_{,\theta}-r^2 V_{,r}^2+V_{,\theta}^2\right]}{r^2}=0
  \,,\label{constraint_3}\\
  & \notag\\
  & F_{0,\theta}\left(F_{1,r}+\frac{1}{r}\right)+F_{1,r} \left(F_{2,\theta}+\cot\theta\right) +F_{1,\theta} \left(F_{2,r}+\frac{1}{r}\right)+F_{0,r}\left(F_{1,\theta}-F_{0,\theta}\right)+F_{2,r}\left(-F_{2,\theta}-\cot \theta\right)
  \notag\\
  & \qquad\qquad
  -F_{2,r\theta}-F_{0,r\theta}+\frac{8 \pi  G e^{-2 F_0} }{r}\left\{H_1 \left[H_2 \left(\omega^2-\mu ^2 e^{2 F_0}\right)+\omega V_{,\theta}\right]+rV_{,r} \left(\omega H_2+V_{,\theta}\right)\right\}=0\,,\label{constraint_4}
  \end{align}
  \end{subequations}
where (\ref{constraint_1}-\ref{constraint_2}) and (\ref{constraint_3}-\ref{constraint_4}) are associated with the scalar and Proca models, respectively.
%%%%%%%%%%%%%%%%%%%%%%%%%%%%%%%%%%%%%%%%%%%
\section{Boundary conditions, physical quantities and numerical methods}
\label{sec:bc}
%%%%%%%%%%%%%%%%%%%%%%%%%%%%%%%%%%%%%%%%%%%
\subsection{Boundary conditions}
%%%%%%%%%%%%%%%%%%%%%%%%%%%%%%%%%%%%%%%%%%%%%%%%%%%%%%%%
In order to construct the SBSs and ABSs, appropriate boundary conditions must be imposed on the ODEs and PDEs we mentioned in Section~\ref{teom}.
In the case of SBSs, the models are constructed according to the following conditions:
\begin{itemize}

\item
At the origin, the matter and metric functions satisfy the following conditions
\begin{eqnarray}
  \partial_r \psi|_{r=0} = 0\,,\quad \partial_r V|_{r=0} = 0 \,,\quad H_1(0) =0 \,,\quad m(0) =0\,,\quad \sigma(r)= \sigma_0\,.
  \end{eqnarray}

\item
We assume that spacetime is asymptotically flat, so the boundary conditions at infinity are
\begin{eqnarray}
  \psi(\infty) = V(\infty)=H_1(\infty)= 0 \,,\quad m(0) =0\,,\quad \sigma(r)= \sigma_0\,.
  \end{eqnarray}

\end{itemize}

%%%%%%%%%%%%%%%%%%%%%%%%%%%%%%%%%%%%%%%%%%%
Additionally, we introduce conditions necessary for constructing the ABSs:
%%%%%%%%%%%%%%%%%%%%%%%%%%%%%%%%%%%%%%%%%%%
\begin{itemize}
  \item  
  The models are subject to the following boundary conditions at the origin
  \begin{eqnarray}
   \partial_r\psi|_{r=0} =V|_{r=0} = H_1|_{r=0} = H_2|_{r=0} = 0 \,,\quad \partial_r F_{0,1,2} = 0\,.
   \end{eqnarray}

  \item 
  All functions vanish at infinity
  \begin{eqnarray}
    \psi|_{r=\infty} = V|_{r=\infty} = H_{1,2}|_{r=\infty} = H_2|_{r=\infty} = 0 \,,\quad F_{0,1,2} = 0\,.
    \end{eqnarray}
  
  \item  On the $z$-axis, regularity and axial symmetry impose the following conditions
  \begin{equation}
  \begin{aligned}
    \partial_\theta \psi|_{\theta=0,\pi} & = \partial_\theta V|_{\theta=0,\pi} = \partial_\theta H_1|_{\theta=0,\pi} = H_2|_{\theta=0,\pi} = 0 \,,\\
     \partial_\theta & F_{0,1,2}|_{\theta=0,\pi} = 0\,,\quad F_1|_{\theta=0,\pi}=F_2|_{\theta=0,\pi}\,.
    \end{aligned} 
  \end{equation} 
  \item
 In this work, the geometry and matter fields of the models we construct are $\mathbb{Z}_2$-even with respect to the equatorial plane, which implies the following conditions at $\theta = \pi/2$:
  \begin{eqnarray}
    \partial_\theta \psi|_{\theta=\frac{\pi}{2}}= \partial_\theta V|_{\theta=\frac{\pi}{2}}= \partial_\theta H_1|_{\theta=\frac{\pi}{2}} = H_2|_{\theta=\frac{\pi}{2}} = 0 \,,\quad \partial_\theta F_{0,1,2} = 0\,.
    \end{eqnarray} 

\end{itemize}

%%%%%%%%%%%%%%%%%%%%%%%%%%%%%%%%%%%%%%%%%%%
\subsection{Physical quantities}
%%%%%%%%%%%%%%%%%%%%%%%%%%%%%%%%%%%%%%%%%%%
The ADM masses $M$ for the SBSs and ABSs can be read off from the asymptotic expansions of metric functions $N(r)$ and $F_0(r,\theta)$
\begin{eqnarray}
  % m(r)= M G + \cdots \,,\quad 
  N(r)= 1-\frac{2M G}{r} + \cdots \,,\quad   
  F_0(r,\theta) = -\frac{MG}{r} + \cdots\,,
  \end{eqnarray} 
the masses can also be obtained from Komar integration:
\begin{eqnarray}
  M^{(s)} = \int \left( T_\mu^{\mu(s)} - 2T_t^{t(s)} \right) \sqrt{-g} drd\theta d\varphi \,,
  \end{eqnarray} 
where the superscript $s = 0,1$ denotes the spin of the matter field.
Additionally, the matter fields $\Psi$ and $\mathcal{A}$ are invariant under a global
$U(1)$ transformation. The corresponding Noether charges are as follows
\begin{eqnarray}
  Q^{(s)} = \int j^{t}_{(s)} \sqrt{-g} drd\theta d\varphi \,,
\end{eqnarray} 
with the conserved currents of scalar and Proca fields:
\begin{eqnarray}
  j^{a}_{(0)} = -i\left( \bar{\Psi} \partial^a \Psi - \Psi \partial^a \bar{\Psi} \right)\,, \qquad
  j^{a}_{(1)} = \frac{i}{2}\left( \bar{\mathcal{F}}^{ab} \mathcal{A}_b - \mathcal{F}^{ab} \bar{\mathcal{A}}_b \right)\,.
\end{eqnarray}

The distribution of matter in space shows a significant difference between the SBSs and ABSs. This difference can be reflected through the \textit{quadrupole moments} of these two models. The quadrupole moment quantifies the extent to which the distribution of matter deviates from spherical symmetry. Therefore, the quadrupole moment of the SBSs is zero. The quadrupole moment of ABSs is encoded in the asymptotic expansions of the metric functions $F_0$ and $F_2$. The asymptotic expansions are \cite{Herdeiro:2015gia}
  \begin{eqnarray}
    \label{quadrupole1}
    F_0(r,\theta)= -\frac{M G}{r} + \frac{f_{03}(\theta)}{r^3} + \cdots \,,\quad  
    F_2(r,\theta) = \frac{M G}{r} + \frac{4a_5+M^2}{4r^2} +\cdots\,,
    \end{eqnarray} 
 with 
 \begin{eqnarray}
  \label{quadrupole2}
  f_{03}(\theta)= b_1 - \frac{12b_1 - 4a_5 M +M^3}{4} \cos^2 \theta  \, ,
  \end{eqnarray} 
where $b_1$ and $a_5$ represent arbitrary constants derived from the numerical results. The quadrupole moment $M_2$ is given by 
\begin{eqnarray}
  \label{quadrupole3}
  M_2 = -\nu_2 - \frac{4}{3}\left( \frac{1}{4} + \frac{B_0}{M^2} \right)M^3 \cos^2 \theta  \,,
  \end{eqnarray} 
with 
\begin{eqnarray}
  \label{quadrupole4}
  B_0 = a_5 - \frac{M^2}{4}\,,\quad \nu_2 = \frac{-12b_1 + 4 a_5 M - M^3}{6}\,.
  \end{eqnarray}
We will analyze the extent to which the ABSs deviate from spherical symmetry and approximate the matter distributions of ABSs by studying their quadrupole moments.

Moreover, the energy densities, $\rho^{(s)} = -T_t^{t(s)}$, more intuitively reflect the characteristics of the matter distributions of the ABSs compared to the quadrupole moments.
The energy density expressions for the scalar and Proca ABSs are as follows:
\begin{subequations}
\begin{align}
& \rho^{(0)} =
 \left( \mu^2 + e^{-2F_0}\omega^2 \right)\psi^2 +
\frac{e^{-2F_1}\left( \psi_{,\theta}^2 + r^2 \psi_{,r}^2 \right)}{r^2}
\,,\\
& \notag\\
& \rho^{(1)} = 
\frac{1}{2} e^{-2F_0 - 2F_1}V_{,r}^2 +
\frac{\omega e^{-2F_0-2F_1} H_1 V_{,r}}{r}+
\frac{e^{-4F_1}H_{2,r}^2}{2r^2}-
\frac{e^{-4F_1}H_{1,\theta}H_{2,r}}{r^3}+
\frac{e^{-2F_0-2F_1}V_{,\theta}^2}{2r^2}+
\notag\\
& \quad \quad \quad 
\frac{\omega e^{-2F_0-2F_1}H_2 V_{,\theta}}{r^2} +
\frac{e^{-4F_1}H_{1,\theta}^2}{2r^4}+
\frac{e^{-2F_1}\left( \mu^2 + \omega^2 e^{-2F_0} \right) \left( H_1^2 + H_2^2 \right)}{2r^2} + 
\frac{\mu^2 e^{-2F_0}V^2}{2}\,.
\end{align}
\end{subequations}
%%%%%%%%%%%%%%%%%%%%%%%%%%%%%%%%%%%%%%%%%%%
\subsection{Numerical methods}\label{methods}
The equations of the models studied in this work possess the following scaling symmetry, which simplifies the numerical calculation process:
\begin{equation}\label{scaling}
  \begin{aligned}
& \psi \rightarrow  \sqrt{4\pi G}\psi 
 \,,\quad
 V \rightarrow \sqrt{4\pi G} V
 \,,\quad
 H_1 \rightarrow \sqrt{4\pi G} H_1
 \,,\quad
 H_2 \rightarrow \sqrt{4\pi G} H_2
 \,,\quad
 r \rightarrow r\mu
 \,,\quad
 \omega \rightarrow \omega/\mu\,,
  \end{aligned}
\end{equation}
where we set $4\pi G = 1$ and $\mu = 1$. 

In the spherical case, the ODEs (\ref{bgequ_spherical}) are solved by using a standard Runge-Kutta ODE solver based on the shooting method, with shooting parameters $\psi(0)$ and $V(0)$ for scalar and Proca cases. While the calculations associated with the PDEs (\ref{bgequ_axial}) are performed with a professional package \textsc{fidisol/cadsol}\cite{Schonauer:1989zwe,SCHONAUER1990279,schoen1}, which uses a Newton-Raphson method. We introduce a new radial coordinate $x$ for the package:
\begin{equation}
  x = \frac{r}{r+c}\,.
\end{equation}
This transformation maps the infinite region $[0,\infty)$ onto a finite interval  $[0,1]$. The constant $c$ can be used to adjust the distribution of the grid within the interval $[0,1]$, therefore, it is necessary to select an appropriate value of $c$ to ensure the accuracy of the numerical results. Additionally, we have tested the accuracy of the numerical solutions by calculating a series of constraints, such as the virial identity, the first-law identity, the equivalence between the ADM mass and the Komar mass, and the constraint equations. The relative error for the solutions reported in this work is less than $10^{-3}$.
%%%%%%%%%%%%%%%%%%%%%%%%%%%%%%%%%%%%%%%%%%%
%%%%%%%%%%%%%%%%%%%%%%%%%%%%%%%%%%%%%%%%%%%
\section{The bifurcations and axisymmetric perturbations}
\label{sec:bif}
%%%%%%%%%%%%%%%%%%%%%%%%%%%%%%%%%%%%%%%%%%%
%%%%%%%%%%%%%%%%%%%%%%%%%%%%%%%%%%%%%%%%%%%
\begin{figure}[h!]
  \centering
  \includegraphics[width=0.49\textwidth]{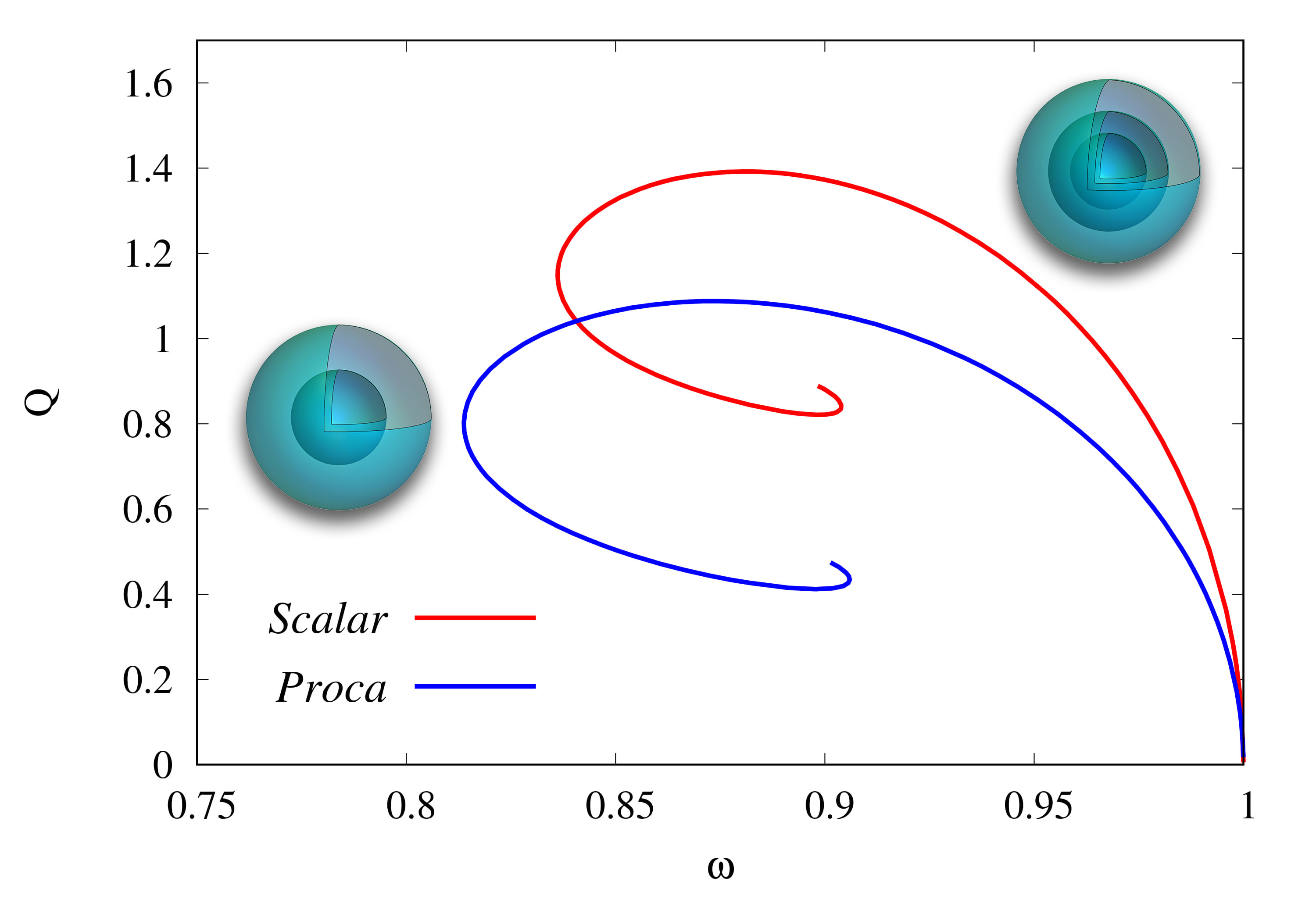}
  \includegraphics[width=0.49\textwidth]{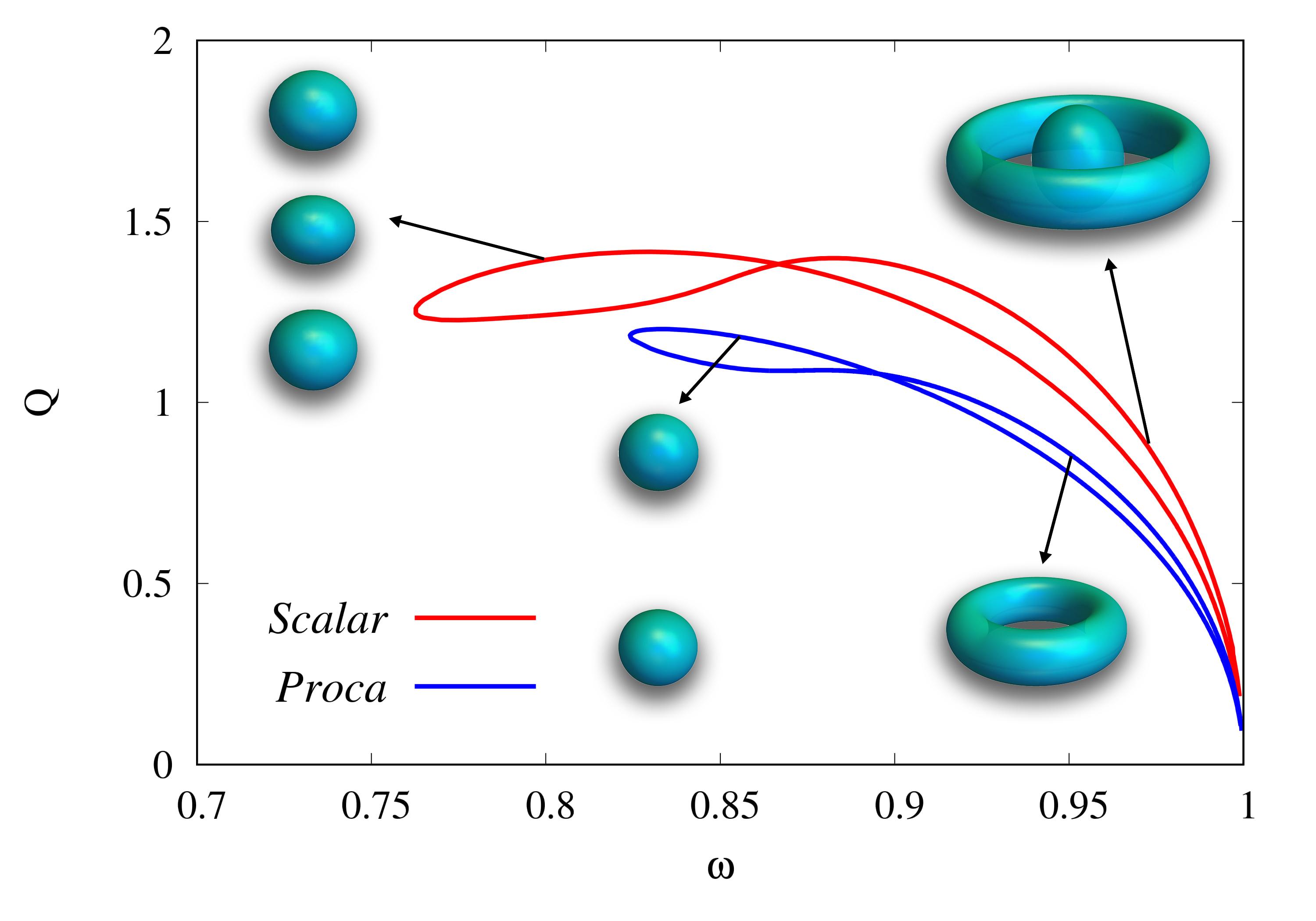}
  \caption{The Noether charges of the ABSs and SBSs as functions of the frequency $\omega$. The morphology of surfaces of constant energy density reflects different matter distributions of the models.}
  \label{q-w}
\end{figure}
Before discussing the bifurcations in BSs, we first delineate the characteristics and differences of the SBSs and ABSs studied in this work. Taking the lines of solutions of BSs depicted in Fig.~\ref{q-w} as examples, the red and blue curves represent scalar and Proca BSs, respectively. In the left panel, the node number $N_S$ of the SBSs is $1$.\footnote{The node number (denoted $N_S$) is the number of zeros of the scalar function (scalar BSs) or of the temporal component of the Proca potential (vector BSs).} The right panel shows the quadrupolar Proca stars proposed in \cite{Herdeiro:2023wqf} and the chains of boson stars with 3 constituents in \cite{Herdeiro:2021mol,Sun:2022duv}. For the Proca and scalar chain-like ABSs, the constituent numbers (denoted $N_A$) are $N_A = 2$ and $N_A = 3$, respectively\footnote{Although $N_A$
is only applicable for categorizing the chain-like ABSs, considering that each chain-like ABSs solution branch always smoothly extends into a ring-like ABSs branch, in this paper we use different values of $N_A$ to represent the entire solution family composed of chain-like ABSs with $N_A$ constituents and the smoothly connected ring-like ABSs.}. Each spherical model's relationship between Noether charge and frequency displays a spiral pattern, with the SBSs becoming more compact as they approach the center of the spiral. In contrast, the situation for ABSs is quite distinct; the curves in the right panel at both ends tend towards the Newtonian limit, forming loop patterns. Additionally, the presence of nodes in the matter function $\psi$ results in a shell around the exterior of scalar SBSs, where the energy density peaks at the origin. The nodes in the function $V$ cause the energy density of Proca SBSs to peak away from the origin, thus the surface of constant energy density forms a spherical shell. 
Remarkably, using the definition of compactness for BSs from the literature, $\mathcal{C} = M/R_{99}$ , where $R_{99}$ is the effective radius containing $99\%$ of the ADM mass $M$ of spacetime, we see that although the maximum ADM mass of SBSs increases with $N_S$, their shell structure means that $R_{99}$ also increases with $N_S$. This causes the maximum compactness, $\mathcal{C}_{max}$, of SBSs with different $N_S$ to be fairly similar. In this work, $\mathcal{C}_{max}$ for SBSs is about $0.12$, staying below the Buchdahl limit of $4/9$ \cite{Buchdahl:1959zz}, with related studies and general results of the Buchdahl limit found in \cite{Andreasson:2007ck,Andreasson:2006ja}.

For the ABSs, a peculiar phenomenon is observed where the matter tends to congregate near the $z$-axis or the equatorial plane, forming chain-like and ring-like structures, respectively. Interestingly, these chain-like and ring-like ABSs can transition smoothly between each other as the frequency $\omega$ changes, despite these forms of matter distribution appearing significantly different. Next, we will analyze the relationship between the SBSs with nodes and the ABSs with unique structures.
%%%%%%%%%%%%%%%%%%%%%%%%%%%%%%%%%%%%%%%%%%%
%%%%%%%%%%%%%%%%%%%%%%%%%%%%%%%%%%%%%%%%%%%
\subsection{The bifurcations in BSs}
%%%%%%%%%%%%%%%%%%%%%%%%%%%%%%%%%%%%%%%%%%%
\begin{figure}[h!]
  \centering
  \includegraphics[width=0.49\textwidth]{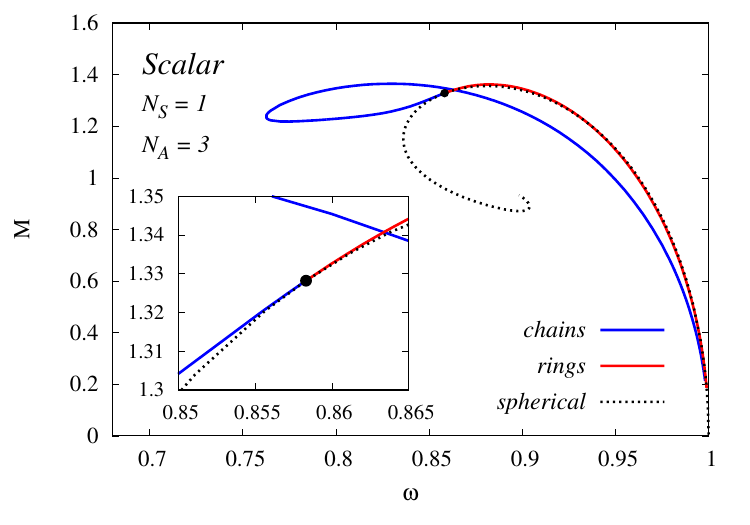}
  \includegraphics[width=0.49\textwidth]{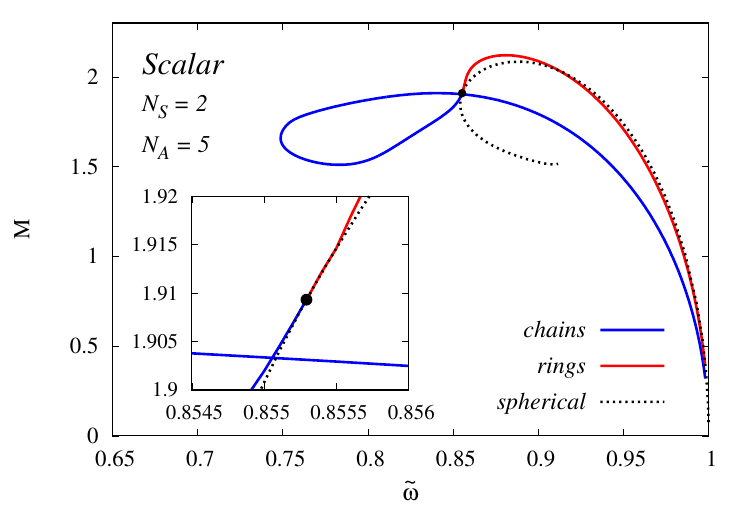}
  \includegraphics[width=0.49\textwidth]{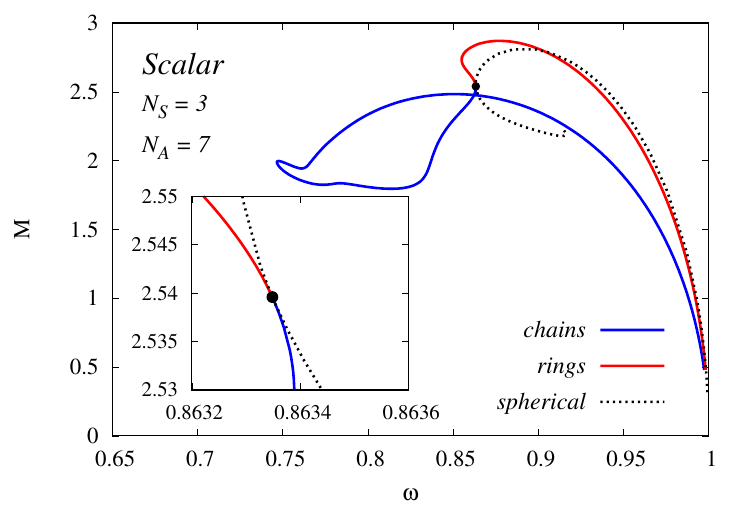}
  \includegraphics[width=0.49\textwidth]{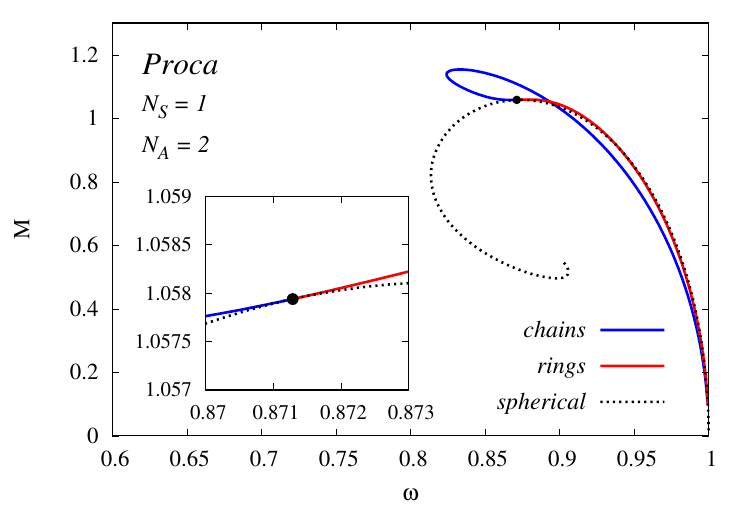}
  \caption{The ADM masses of the ABSs and SBSs as functions of the frequency $\omega$. The black points indicate bifurcations.}
  \label{m-q-w}
\end{figure}
\begin{table}[h!]
  \centering 
  \begin{tabular}{|c|c|c|c|c|c|}
  \hline
  Spin-$s$ & $N_S$ & $N_A$ & $\omega$ & $M$ & $Q$\\
  \hline
  \multirow{3}{*}{$s=0$} & 1 & 3 & 0.8583 & 1.3282 & 1.3597\\
  \cline{2-6}
  & 2 & 5 & 0.8553 & 1.9099 & 1.9346\\
  \cline{2-6}
  & 3 & 7 & 0.8634 & 2.5388 & 2.5663\\
  \hline
  $s=1$ & 1 & 2 & 0.8713 & 1.0579 & 1.0879 \\
  \hline
  \end{tabular}
  \caption{The parameters and physical quantities at several bifurcation points shown in Fig.~\ref{m-q-w}}
  \label{parameters}
  \end{table}
By comparing various physical quantities of the SBSs and ABSs, we find a connection, and transitions, between BSs with different symmetries. As the frequency changes, both the chain-like and ring-like ABSs branches can smoothly connect to the branches of the SBSs. In other words, two branches of chain-like and ring-like ABSs can bifurcate from the SBSs trunk, as shown in Fig.~\ref{m-q-w}. For the scalar case, SBSs trunks with $N_S = 1,2,3$ can bifurcate into ABSs branches with $N_A = 3,5,7$ respectively. We propose that this type of bifurcation exists between scalar SBSs and ABSs under the following empirical condition:
\begin{eqnarray}\label{nsna}
  N_S = \frac{N_A - 1}{2}\,.
  \end{eqnarray}
However, in this work, we do not construct cases where $N_S > 3$ (and $N_A > 7$) due to numerical challenges. For the Proca case, we only find bifurcations between the SBSs with $N_S = 1$ and the ABSs with $N_A = 2$, but we do not rule out the possibility of bifurcations for the Proca SBSs with $N_S > 1$. It is important to note that although the red curves and the black dashed lines in the figure seem to overlap in some parts, we will later see that these two models, the SBSs and the ring-like ABSs, exhibit significant differences. Table~\ref{parameters} presents the frequencies, ADM mass, and Noether charge at these bifurcation points. The physical quantities of the SBSs and ABSs are identical at the bifurcation points. Next, we will explain this type of bifurcation in BSs from a perturbative perspective.

%%%%%%%%%%%%%%%%%%%%%%%%%%%%%%%%%%%%%%%%%%%
\subsection{Axisymmetric perturbation}
%%%%%%%%%%%%%%%%%%%%%%%%%%%%%%%%%%%%%%%%%%%
We consider axisymmetric perturbations as follows:
\begin{equation}\label{ansatz_pert}
  \begin{aligned}
    \bar{g}_{tt} & =- N(r) \sigma^2(r)(1 - \epsilon H(r)P_\ell(\cos \theta)), &\quad   \bar{\psi}(r,\theta) & = \psi(r) + \epsilon \psi_1(r)P_\ell(\cos \theta), \\
    \bar{g}_{rr} & = \frac{1}{N(r)}(1 + \epsilon H(r)P_\ell(\cos \theta)), &\quad   \bar{V}(r,\theta) & = V(r) + \epsilon A_0(r)P_\ell(\cos \theta), \\
    % \bar{g}_{tt} & =- \left(1-\frac{2m(r)}{r} \right) \sigma^2(r)(1 - \epsilon H(r)P_\ell(\cos \theta)), &\quad   \bar{\psi}(r,\theta) & = \psi(r) + \epsilon \psi_1(r)P_\ell(\cos \theta), \\
    % \bar{g}_{rr} & = \frac{1}{1-\frac{2m(r)}{r}}(1 + \epsilon H(r)P_\ell(\cos \theta)), &\quad   \bar{V}(r,\theta) & = V(r) + \epsilon A_0(r)P_\ell(\cos \theta), \\
    \bar{g}_{\theta \theta} & = r^2 (1 + \epsilon K(r)P_\ell(\cos \theta)), &\quad   \bar{H}_1(r,\theta) & = H_1(r) + \epsilon A_1(r)P_\ell(\cos \theta), \\
    \bar{g}_{\varphi \varphi} & = r^2 \sin^2 \theta (1 + \epsilon K(r)P_\ell(\cos \theta)), &\quad   \bar{H}_2(r,\theta) & = \epsilon A_2(r)\frac{P_\ell(\cos \theta)}{d\theta}\, ,
  \end{aligned}
  \end{equation}
where $\epsilon$ is a small parameter, $P_\ell(\cos \theta)$ is the Legendre polynomial. $H(r)$, $K(r)$, $\psi_1(r)$, $A_0(r)$, $A_1(r)$, $A_2(r)$ are perturbation functions. Different values of $\ell$ correspond to different modes of the axisymmetric perturbations. We do not conduct a systematic analysis of the stability of SBSs under the axisymmetric perturbations. Instead, we focus only on the critical points of stability under the perturbations, the zero-modes. Therefore, the perturbations (\ref{ansatz_pert}) are independent of time.

Substituting (\ref{ansatz_pert}) into (\ref{E-eq}) and (\ref{KG-eq}) and taking only the first order terms in $\epsilon$, we can derive a set of perturbation equations satisfied by the perturbation functions. Subsequently, using the $(r,r)$, $(r,\theta)$ and $(\theta,\theta)$ Einstein equations, the functions $K(r)$, $K'(r)$, and $K''(r)$ can be expressed in terms of the other functions. Finally, these perturbation equations can be simplified into two (three) second order linear ODEs for the scalar (Proca) cases

\begin{equation}\label{p-eq-scalar}
  \left. \text{Scalar}\hspace{1em} \middle\vert
  \begin{aligned}
    H''(r) + 
    a_1 H'(r)+
    a_2 H(r)+
    a_3 \psi_1(r) & = 0\,,\\
    \psi_1''(r) + 
    b_1\psi_1'(r)+
    b_2\psi_1(r)+
    b_3 H(r) & = 0\, ,
  \end{aligned}
  \right.
\end{equation}
\\
\begin{equation}\label{p-eq-proca}
  \left. \text{Proca}\hspace{1em} \middle\vert
  \begin{aligned}
    H''(r) + 
    c_1 H'(r)+
    c_2 H(r)+
    \mathcal{C}(r) & = 0\,,\\
    A_0''(r) + 
    d_1 A_0'(r)+
    d_2 A_2'(r)+
    d_3 H'(r)+
    d_4 A_0(r)+
    d_5 A_2(r)+
    d_6 H(r) & = 0\,,\\
    A_2''(r) + 
    e_1 A_0'(r)+
    e_2 A_2'(r)+
    e_3 H'(r)+
    e_4 A_0(r)+
    e_5 A_2(r)+
    e_6 H(r) & = 0\, ,
  \end{aligned}
  \right.
\end{equation}
where $\mathcal{C}(r)$ in the first equation of \ref{p-eq-proca} is related to $A_0(r)$ and $A_2(r)$:
\begin{align}
\mathcal{C} & = \frac{4 A_0 \kappa  \left(H_1 N \sigma ^2 \left(-H_1^2 \kappa  N^2 r^2 \sigma ^2+N \omega ^2 \left(H_1^2 \kappa 
   r^2-1\right)+\omega ^2\right)+H_1 \kappa  r^2 V^2 \omega ^2+2 r V \omega ^3\right)}{N^2 r \sigma ^2 \omega ^3} +\notag\\
      &\quad \frac{4 H_1 \kappa  N \sigma ^2 \left(A_0' \left(\ell  (\ell +1)+r^2\right)-\ell  (\ell +1) \omega 
      A_2'\right)}{r^2 \omega ^3-N \sigma ^2 \omega  \left(\ell  (\ell +1)+r^2\right)}\, .
\end{align}
In the Proca case, the equation for $A_1$ is an algebraic equation, which can be expressed as 
\begin{align}
  A_1 = \frac{r^2\omega A_0' - \sigma^2 N \left( r^2 \mu^2 H H_1 + 6A_2' \right)}{r^2 \left( \omega^2 - \mu^2 \sigma^2 N \right)-6 \sigma^2 N}\,.
\end{align}
 The coefficients in (\ref{p-eq-scalar}) and (\ref{p-eq-proca}) are defined by the background solutions. The coefficients for scalar case are as follows:
\begin{subequations}
  \begin{align}
    a_1 & =  \frac{1}{r} + \frac{1}{rN} + \frac{2\kappa \mu^2 r \psi^2}{N}\,, \\
  a_2 & = - \frac{1}{r^2} \left( 1 + \frac{1}{N^2} + \frac{\ell(\ell+1)-2}{N} \right)
  - 4\kappa^2 r^2 \psi'^4 +
  4\kappa \psi'^2 \left( 1 - \frac{1}{N} \right) + \notag\\
      &\qquad \frac{4\kappa \psi^2}{N}\left( 
  -\mu^2 + \frac{1}{N} \left( \mu^2 + \frac{5 \omega^2}{\sigma^2}\right)
   + 2\kappa \psi'^2 r^2 \left(\mu^2 - \frac{\omega^2}{N\sigma^2} \right) - \frac{\omega^2}{N^2\sigma^2}
   \right)+ \notag\\
   &\qquad r^2 \phi^4 \frac{4 \kappa^2}{N^2} \left( -\mu^4 - \frac{\omega^4}{N^2 \sigma^4} + \frac{2 \omega^2 \mu^2}{N\sigma^2} \right)\,,\\
  a_3 & = 16 \kappa^2 r \psi'^3 + 
  \frac{8\kappa \psi'}{r} \left( \frac{1}{N} - 1 \right) + 
  \frac{8\kappa \psi}{N} \left( \frac{2\omega^2}{N\sigma^2}  \right)+
  \frac{16\kappa^2 r\psi^2 \psi'}{N}\left( \frac{\omega^2}{N\sigma^2} - \mu^2\right)\,,\\
  b_1 & =  \frac{1}{r} + \frac{1}{rN} + \frac{2\kappa \mu^2 r \psi^2}{N}\,, \\
  b_2 & = -\frac{\ell(\ell+1)}{Nr^2} - 8\kappa \psi'^2 -\frac{\mu^2}{N} + \frac{\omega^2}{N^2\sigma^2} \,,\\
  b_3 & = \frac{\psi'}{r}\left( \frac{1}{N} - 1 \right)+
  2 \kappa r \psi'^3   +
  \frac{\psi}{N}\left( \frac{2\omega^2}{N\sigma^2} - \mu^2 \right) +
  \frac{2\kappa r \psi^2 \psi'}{N}\left( \frac{\omega^2}{N\sigma^2} - \mu^2 \right)  \,,
  \end{align}
\end{subequations}
whereas the coefficients for the Proca case are more complicated:
\begin{subequations}
  \begin{align}
  c_1 & = -\frac{H_1^2 \kappa  N r \sigma ^2}{\omega ^2}+\frac{1}{N r}+\frac{1}{r}\,, \\
  c_2 & = -\frac{2 \kappa  V^2 \left(-H_1^2 \kappa  N^2 r^2 \sigma ^2+N \omega ^2 \left(H_1^2 \kappa  r^2-5\right)+\omega
   ^2\right)}{N^3 \sigma ^2 \omega ^2}-\frac{\kappa ^2 r^2 V^4}{N^4 \sigma ^4} -\notag\\
      &\quad \frac{H_1^4 \kappa ^2 N^5 r^4 \sigma ^6 \left(\ell  (\ell +1)+r^2\right)-H_1^4 \kappa ^2 N^4 r^4 \sigma ^4 \omega ^2
      \left(2 \ell  (\ell +1)+3 r^2\right)-r^2 \omega ^6}{N^3 r^2 \sigma ^2 \omega ^4 \left(\ell  (\ell
      +1)+r^2\right)-N^2 r^4 \omega ^6} -\notag\\
      &\quad \frac{H_1^2 (-\kappa ) r^4 \omega ^2 \left(H_1^2 \kappa  N r^2+2\right)+\sigma ^2 \left(\ell  (\ell
      +1)+r^2\right)+(2-\ell  (\ell +1)) r^2 \omega ^2}{N^2 r^2 \sigma ^2 \left(\ell  (\ell +1)+r^2\right)-N r^4
      \omega ^2} -\notag\\
      &\quad \frac{2 H_1^2 \kappa  r^2 \left(\ell  (\ell +1) \sigma ^2+r^2 \omega ^2+2 r^2\right)+(\ell  (\ell +1)-2) \sigma ^2
      \left(\ell  (\ell +1)+r^2\right)-r^2 \omega ^2}{N r^2 \sigma ^2 \left(\ell  (\ell +1)+r^2\right)-r^4 \omega ^2} -\notag\\
      &\quad \frac{H_1^4 \kappa ^2 r^2 \left(\ell  (\ell +1)+3 r^2\right)-2 H_1^2 \kappa  \left(\ell  (\ell +1)+3 r^2\right)}{N
      \left(\ell  (\ell +1)+r^2\right)-r^4 \omega ^2} -\notag\\
      &\quad \frac{\omega ^2 \left(\ell  (\ell +1)+r^2\right)-2 H_1^2 \kappa  r^2 \sigma ^2 \left(\ell  (\ell +1)+r^2\right)}{N
      r^2 \omega ^2 \left(\ell  (\ell +1)+r^2\right)-r^4 \omega ^4}\,,\\
  d_1 & = \frac{2 N^2 \sigma ^2 \left(-\ell  (\ell +1)+\frac{1}{2} \ell  (\ell +1) H_1^2 \kappa  r^2-r^2\right)+N r^2 \omega
  ^2+\kappa  r^2 V^2 \left(\ell  (\ell +1)+r^2\right)+r^2 \omega ^2}{N r^3 \omega ^2-N^2 r \sigma ^2 \left(\ell 
  (\ell +1)+r^2\right)}\,, \\
  d_2 & = \frac{\ell  (\ell +1) \left(N \sigma ^2 \left(-H_1^2 \kappa  N^2 r^2 \sigma ^2+N \omega ^2 \left(H_1^2 \kappa 
  r^2-1\right)+\omega ^2\right)+\kappa  r^2 V^2 \omega ^2\right)}{N^2 r \sigma ^2 \omega  \left(\ell  (\ell
  +1)+r^2\right)-N r^3 \omega ^3}\,, \\
  d_3 & = -\frac{N \sigma^2 H_1}{\omega}\,, \\
  d_4 & = 4 H_1^2 \kappa  \left(\frac{N \sigma ^2}{\omega ^2}-1\right)+\frac{\omega ^2}{N^2 \sigma ^2}+\frac{1}{N}\left(\frac{\ell  (\ell
  +1)}{r^2}+1\right)\,,\\
  d_5 & = 0\,,\\
  d_6 & = \frac{r V \left(2 \omega ^2-N \sigma ^2\right) \left(r^2 \omega ^2-N \sigma ^2 \left(\ell  (\ell
  +1)+r^2\right)\right)}{N^2 \sigma ^2 \left(r^3 \omega ^2-N r \sigma ^2 \left(\ell  (\ell +1)+r^2\right)\right)}+ \notag\\
      &\quad \frac{H_1 \kappa  r^2 V^2 \left(N^2 \sigma ^4 \left(\ell  (\ell +1)+r^2\right)-N \sigma ^2 \omega ^2 \left(\ell 
      (\ell +1)+3 r^2\right)+r^2 \omega ^4\right)}{N^2 \sigma ^2 \omega  \left(r^3 \omega ^2-N r \sigma ^2 \left(\ell 
      (\ell +1)+r^2\right)\right)}+\notag\\
      &\quad \frac{H_1 \left(N^2 \sigma ^4 \left(\ell  (\ell +1)+r^2\right)-N \sigma ^2 \omega ^2 \left(\ell  (\ell +1)+3
      r^2\right)+r^2 \omega ^4\right) \left(-H_1^2 \kappa  N^2 r^2 \sigma ^2+N \omega ^2 \left(H_1^2 \kappa 
      r^2-1\right)+\omega ^2\right)}{N \omega  \left(r^3 \omega ^2-N r \sigma ^2 \left(\ell  (\ell
      +1)+r^2\right)\right)}\,,\\
  e_1 & = \frac{2 r \omega }{r^2 \omega ^2-N \sigma ^2 \left(\ell  (\ell +1)+r^2\right)}\,,\\
  e_2 & = \frac{-H_1^2 \kappa N^3 r^2 \sigma ^4 \left(\ell  (\ell +1)+r^2\right)+N \left(\sigma ^2 \omega ^2
  \left(\ell  (\ell +1)+r^2\right)+r^2 \omega ^4\right)-r^2 \omega ^4}{N r \omega ^2 \left(N \sigma ^2 \left(\ell 
  (\ell +1)+r^2\right)-r^2 \omega ^2\right)}+\notag\\
      &\qquad \frac{N \sigma ^2 \left(\ell  (\ell +1)+H_1^2 \kappa  r^4-r^2\right)}{r \left(N \sigma ^2 \left(\ell  (\ell
      +1)+r^2\right)-r^2 \omega ^2\right)}\,,\\
  e_3 & = 0\,,\\
  e_4 & = -\frac{4 H_1^2 \kappa }{\omega }\,,\\
  e_5 & = \frac{\omega ^2}{N^2 \sigma ^2}-\frac{1}{N} \left(\frac{\ell  (\ell +1)}{r^2}+1\right)\,,\\
  e_6 & = \frac{H_1 \kappa  r^2 V^2 \left(N \sigma ^2 \left(\ell  (\ell +1)+r^2\right)-r^2 \omega ^2\right)+V \left(2 N r
  \sigma ^2 \omega  \left(\ell  (\ell +1)+r^2\right)-2 r^3 \omega ^3\right)}{N^2 \sigma ^2 \left(N r \sigma ^2
  \left(\ell  (\ell +1)+r^2\right)-r^3 \omega ^2\right)}+ \notag\\
      &\quad \frac{H_1 \left(H_1^2 \kappa  \left(-N^3\right) r^2 \sigma ^4 \left(\ell  (\ell +1)+r^2\right)+N^2 \sigma ^2 \omega
      ^2 \left(-\ell  (\ell +1)+2 H_1^2 \kappa  r^2 \left(\frac{1}{2} \ell  (\ell +1)+r^2\right)+r^2\right)-r^2
      \omega ^4\right)}{N \omega ^2 \left(N r \sigma ^2 \left(\ell  (\ell +1)+r^2\right)-r^3 \omega ^2\right)} + \notag\\
      &\quad \frac{H_1 \left(H_1^2 (-\kappa ) r^4 \omega ^2+\sigma ^2 \left(\ell  (\ell +1)+r^2\right)+r^2 \omega ^2\right)}{N r
      \sigma ^2 \left(\ell  (\ell +1)+r^2\right)-r^3 \omega ^2}\,,
\end{align}
\end{subequations}
where $\kappa = 4\pi G$. Following the setting in Section~\ref{methods}, we set $\kappa = 1$. The power expansions of the perturbation functions are shown as follows:
\begin{equation}\label{p-series-scalar}
  \left. \text{Scalar}\hspace{1em} \middle\vert
  \begin{aligned}
    H(r)= h_\ell r^\ell + \mathcal{O}(r^{\ell+2})\,, \qquad 
    \psi_1(r)= p_\ell r^\ell + \mathcal{O}(r^{\ell+2})\,,\, 
  \end{aligned}
  \right.
\end{equation}
\begin{equation}\label{p-series-proca}
  \left. \text{Proca}\hspace{1em} \middle\vert
  \begin{aligned}
    H(r)= h_\ell r^\ell + \mathcal{O}(r^{\ell+2})\,, \qquad 
    A_0(r)= \alpha_\ell r^\ell + \mathcal{O}(r^{\ell+2})\,,
    \qquad
    A_2(r)= \beta_\ell r^\ell + \mathcal{O}(r^{\ell+2})\, .
  \end{aligned}
  \right.
\end{equation}
Obviously, we can exploit the linearity of the perturbation equations, (\ref{p-eq-scalar}) and (\ref{p-eq-proca}), to set $h_l=1$, thus simplifying the process of solving these two sets of equations.
This adjustment allows us to reduce the study of zero-modes in the scalar and Proca cases. The perturbation functions only satisfy the asymptotically flat conditions $H(\infty)=0$, $\psi_1(\infty)=0$, $A_0(\infty)=0$ and $A_2(\infty)=0$ when the background SBSs correspond to the zero-modes. Without confirming whether the bifurcation points shown previously coincide with the zero-modes of SBSs, we employ a multi-parameter shooting method to solve (\ref{p-eq-scalar}) and (\ref{p-eq-proca}), aiming to identify the SBSs background corresponding to the zero-modes. The shooting parameters are $\omega$, $h_{\ell}$, $p_{\ell}$ and $\omega$, $h_{\ell}$, $\alpha_{\ell}$, $\beta_{\ell}$ for scalar and Proca cases, respectively.

%%%%%%%%%%%%%%%%%%%%%%%%%%%%%%%%%%%%%%%%%%%
\section{The numerical results for perturbations}\label{sec:result}
%%%%%%%%%%%%%%%%%%%%%%%%%%%%%%%%%%%%%%%%%%%
Our objective is to study the relationship between the SBSs and the ABSs through axisymmetric perturbations. Considering that the SBSs have even-parity, and all ABSs constructed in this work also possess even-parity, the modes of axisymmetric perturbations must be consistent with the parities of these BSs, meaning $\ell$ can only be even numbers. In this section, we discuss the perturbation modes for $\ell = 2$ and $\ell = 4$.
%%%%%%%%%%%%%%%%%%%%%%%%%%%%%%%%%%%%%%%%%%%
\subsection{The results for $\ell=2$}
%%%%%%%%%%%%%%%%%%%%%%%%%%%%%%%%%%%%%%%%%%%
\begin{figure}[h!]
  \centering
  \includegraphics[width=0.49\textwidth]{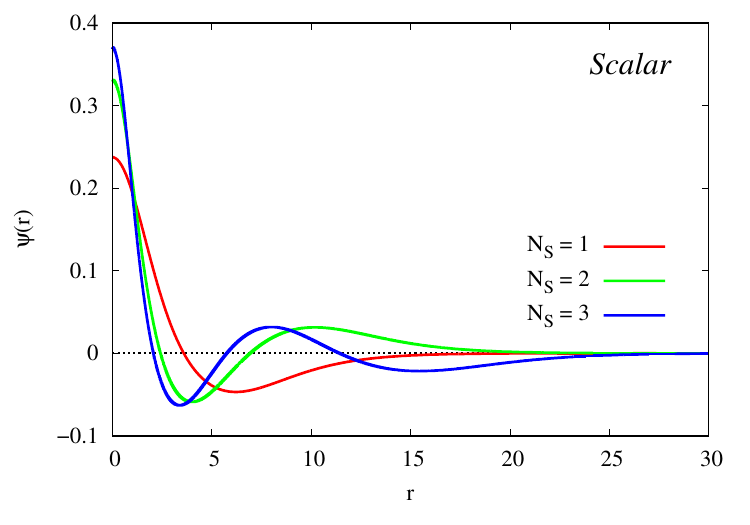}
  \includegraphics[width=0.49\textwidth]{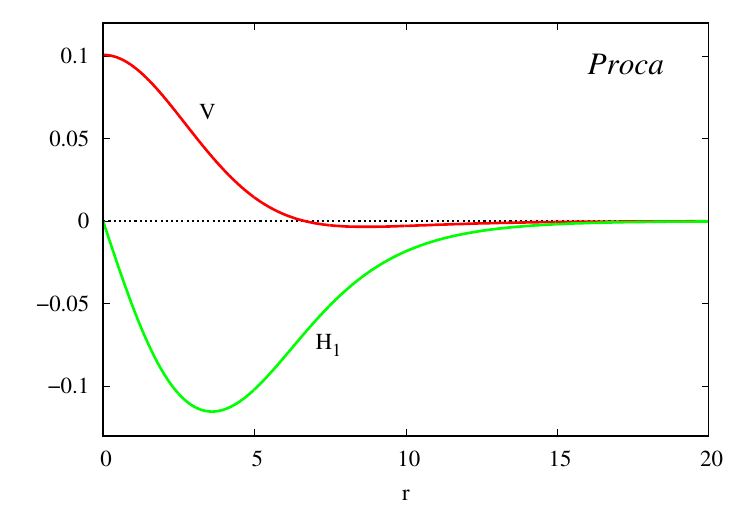}
  \caption{The matter field profiles of the SBSs backgrounds related to the zero-modes for scalar and Proca cases.}
  \label{back-matter}
\end{figure}
By solving the perturbation equations (\ref{p-eq-scalar}) and (\ref{p-eq-proca}) using the multi-parameter shooting method, we find that for each type of SBSs we study in this work, there is always a unique background solution that satisfies the condition where all perturbation functions vanish asymptotically, and this SBS background corresponds to the zero-mode. The matter field profiles of the SBSs backgrounds related to the zero-modes are shown in Fig.~\ref{back-matter}. The profiles in the left panel show one (red curve), two (green curve), and three (blue curve) nodes, which correspond to the first three excited states of scalar SBSs, respectively. The profiles in the right panel correspond to the lowest state of Proca SBSs\footnote{The ground state of the Proca stars was found in~\cite{Herdeiro:2023wqf}.}, where $V(r)$ has one node and $H_1$ has no nodes. After comparing the relevant parameters, we can confirm the consistency between the SBS backgrounds in Fig.~\ref{back-matter} and the SBSs at the bifurcation points in Fig.~\ref{m-q-w}. This implies that the bifurcation points in BSs are the critical points of stability for SBSs under $\ell = 2$ axisymmetric perturbations, where the SBSs bifurcate into chain-like and ring-like ABSs.
\begin{figure}[h!]
  \centering
  \includegraphics[width=0.49\textwidth]{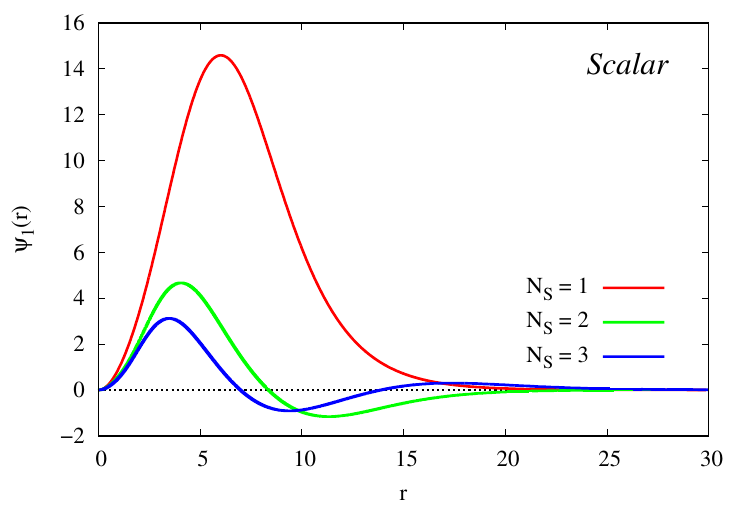}
  \includegraphics[width=0.49\textwidth]{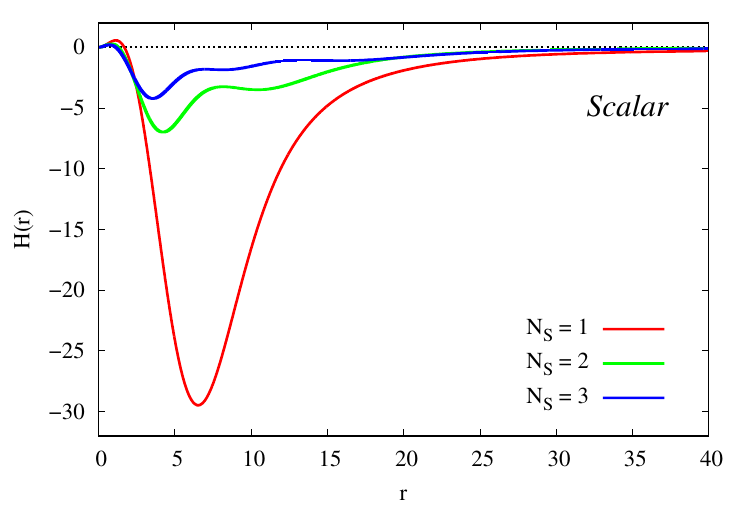}
  \includegraphics[width=0.49\textwidth]{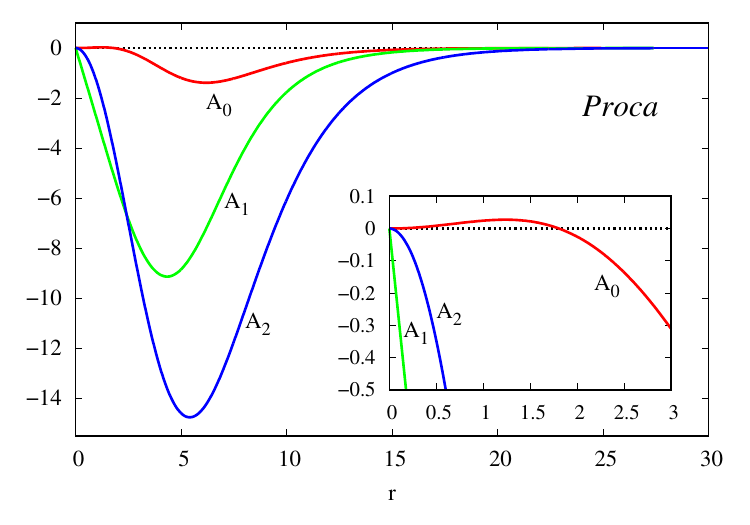}
  \includegraphics[width=0.49\textwidth]{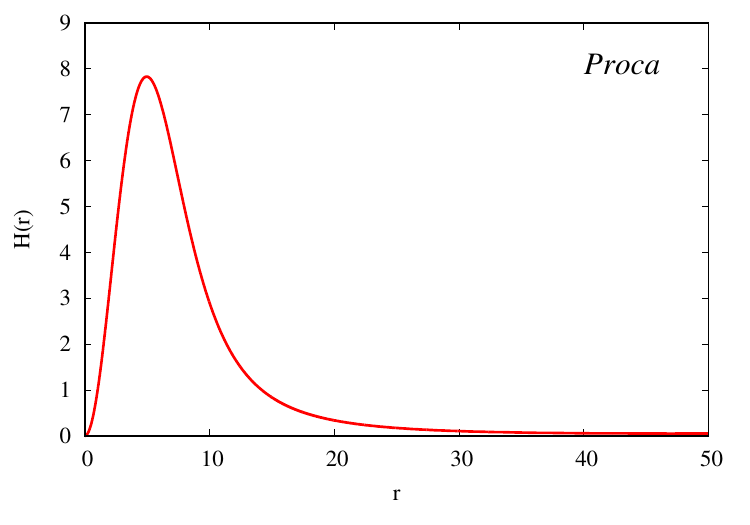}
  \caption{The profiles of perturbation functions $\psi_1(r)$, $A_0(r)$, $A_1(r)$, $A_2(r)$ (left column) and $H(r)$ (right column) at the bifurcation points for scalar (top row) and Proca (bottom row) cases.}
  \label{h_psi1}
\end{figure}
Additionally, the profiles of the perturbation functions at bifurcation points in scalar and Proca cases are shown in Fig.~\ref{h_psi1}.
As expected, all perturbation functions vanish at infinity. The node number of the perturbation function $\Psi_1(r)$ is always one less than $N_s$, while those of $A_0$ and $A_1$ are consistent with the background solutions. 

\begin{table}[h!]
  \centering
  \begin{tabular}{|>{\centering\arraybackslash}m{0.4cm}|>{\centering\arraybackslash}m{1.7cm}|>{\centering\arraybackslash}m{1.7cm}|>{\centering\arraybackslash}m{1.7cm}|>{\centering\arraybackslash}m{1.7cm}|>{\centering\arraybackslash}m{1.7cm}|>{\centering\arraybackslash}m{1.7cm}|>{\centering\arraybackslash}m{1.7cm}|}
      \hline
      $N_S$\vspace{0.05cm} & 
      Chains\vspace{0.05cm} & 
      \multicolumn{2}{>{\centering\arraybackslash}m{3.4cm}|}{\vspace{0.3cm}\hspace*{0cm}$\overset{\text{\large $\epsilon < 0$}}{\xleftarrow{\hspace{2.8cm}}}$\hspace*{1cm}\vspace{0cm}} & 
      Spherical\vspace{0.05cm}&
      \multicolumn{2}{>{\centering\arraybackslash}m{3.4cm}|}{\vspace{0.3cm}\hspace*{0cm}$\overset{\text{\large $\epsilon > 0$}}{\xrightarrow{\hspace{2.8cm}}}$\hspace*{1cm}\vspace{0cm}} & 
      Rings\vspace{0.05cm} \\ \hline
      1 & \vspace{0.2cm}\includegraphics[height=1.5cm, keepaspectratio]{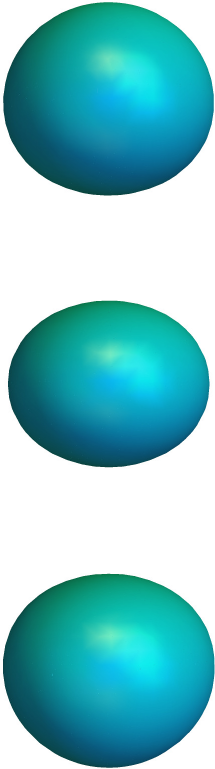} & \vspace{0.2cm}\includegraphics[height=1.5cm, keepaspectratio]{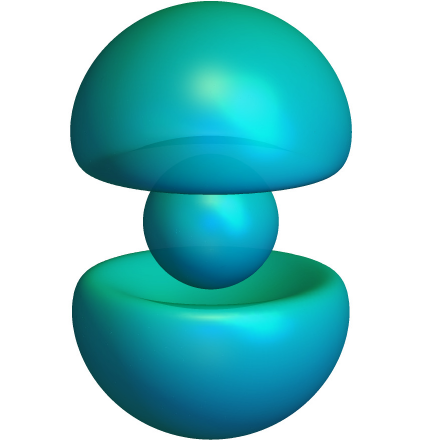} & \vspace{0.2cm}\includegraphics[height=1.5cm, keepaspectratio]{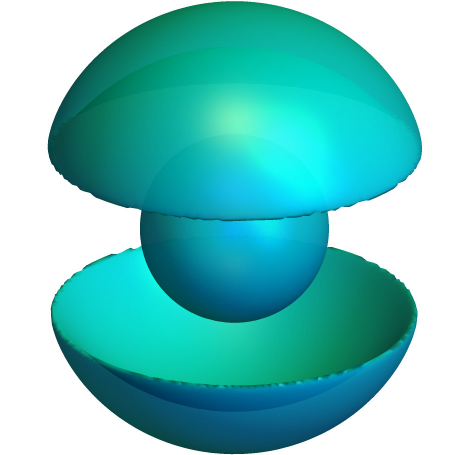} & \vspace{0.2cm}\includegraphics[height=1.5cm, keepaspectratio]{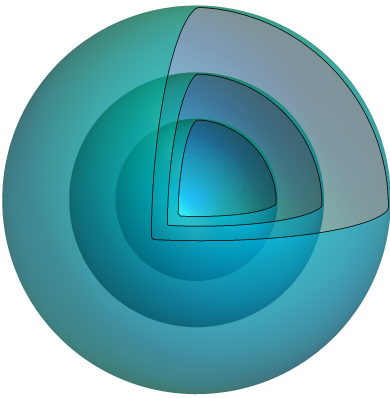} & \vspace{0.2cm}\includegraphics[height=1.5cm, keepaspectratio]{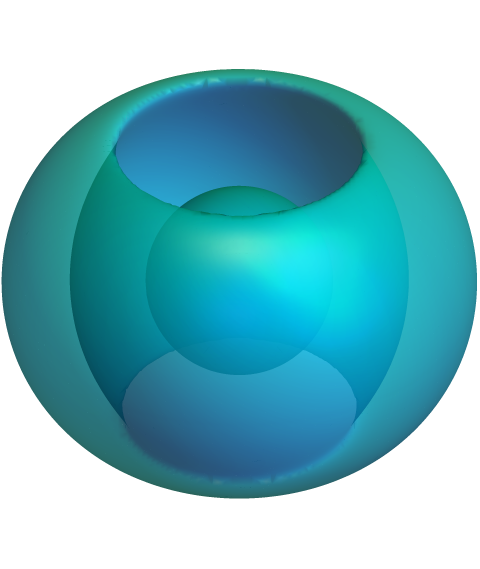} & \vspace{0.2cm}\includegraphics[height=1.5cm, keepaspectratio]{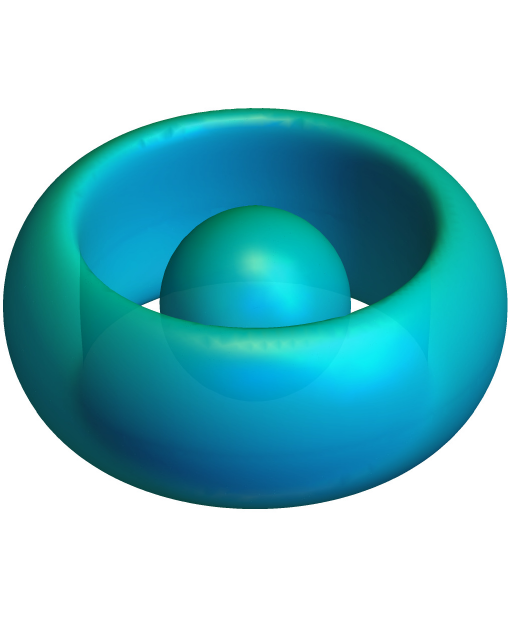} & \vspace{0.2cm}\includegraphics[height=1.5cm, keepaspectratio]{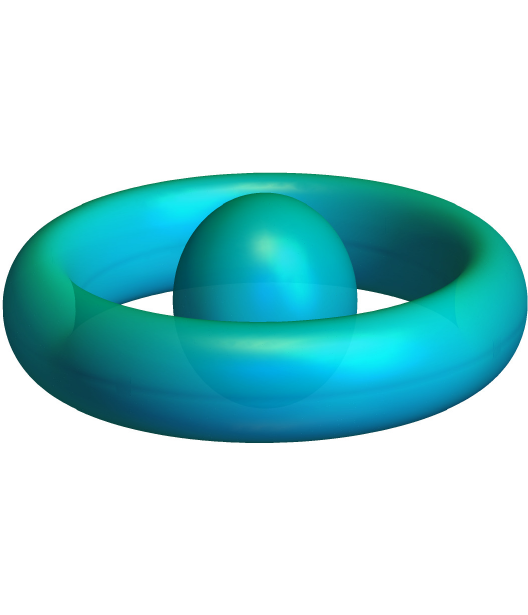} \\ \hline
      2 & \vspace{0.2cm}\includegraphics[height=1.5cm, keepaspectratio]{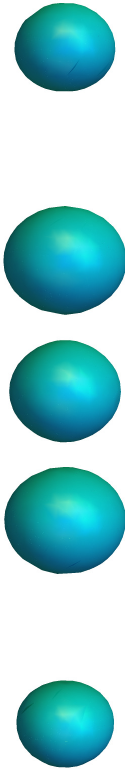} & \vspace{0.2cm}\includegraphics[height=1.5cm, keepaspectratio]{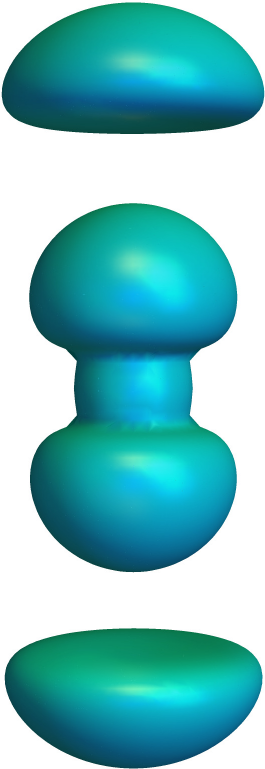} & \vspace{0.2cm}\includegraphics[height=1.5cm, keepaspectratio]{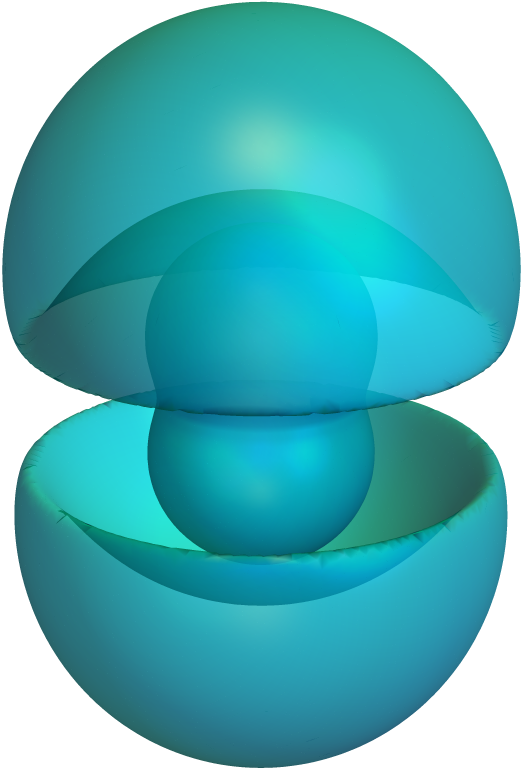} & \vspace{0.2cm}\includegraphics[height=1.5cm, keepaspectratio]{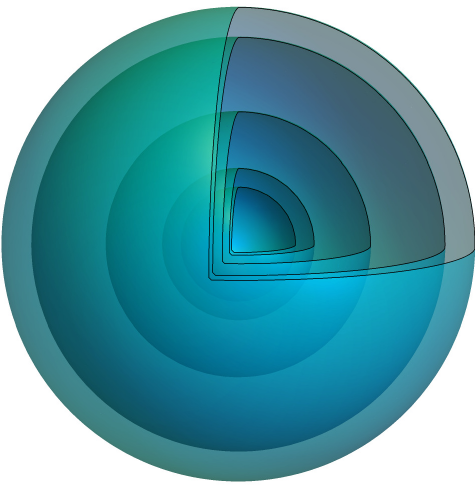} & \vspace{0.2cm}\includegraphics[height=1.5cm, keepaspectratio]{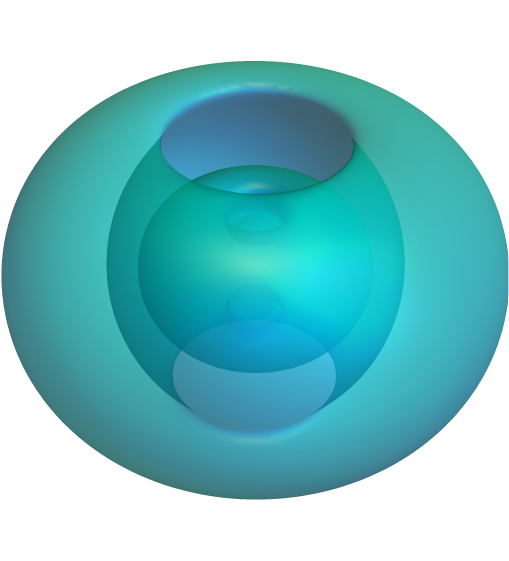} & \vspace{0.2cm}\includegraphics[height=1.5cm, keepaspectratio]{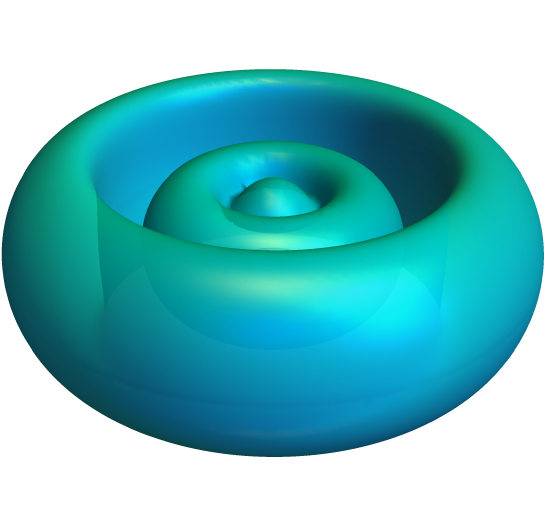} & \vspace{0.2cm}\includegraphics[height=1.5cm, keepaspectratio]{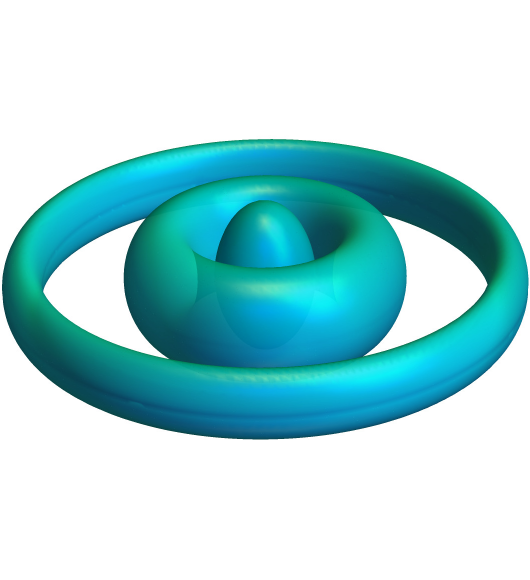} \\ \hline
      3 & \vspace{0.2cm}\includegraphics[height=1.5cm, keepaspectratio]{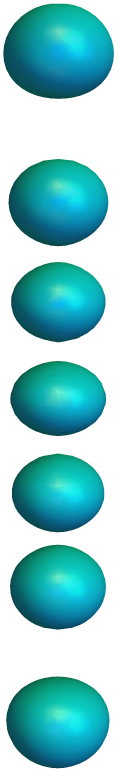} & \vspace{0.2cm}\includegraphics[height=1.5cm, keepaspectratio]{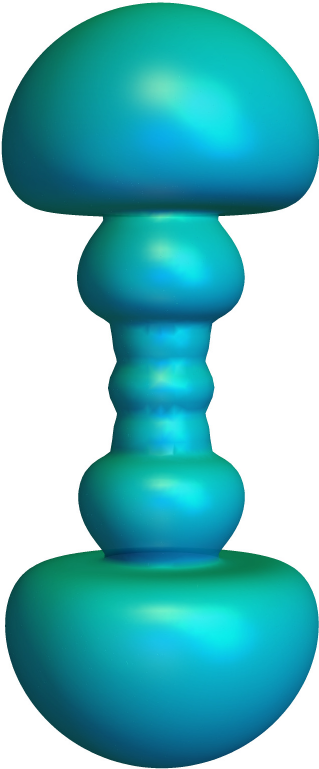} & \vspace{0.2cm}\includegraphics[height=1.5cm, keepaspectratio]{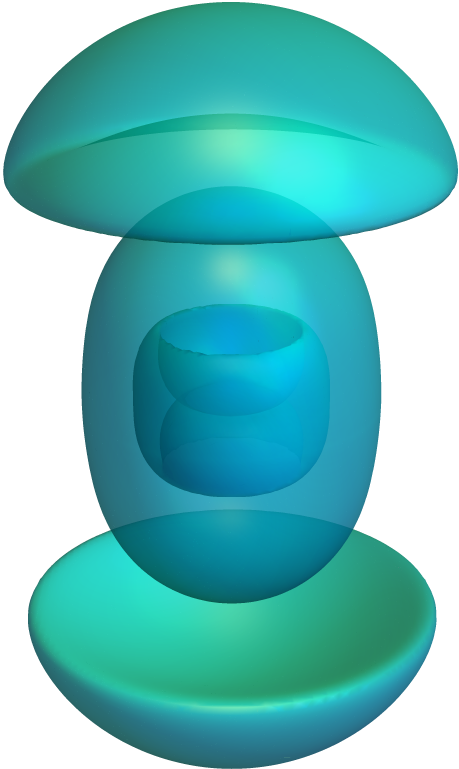} & \vspace{0.2cm}\includegraphics[height=1.5cm, keepaspectratio]{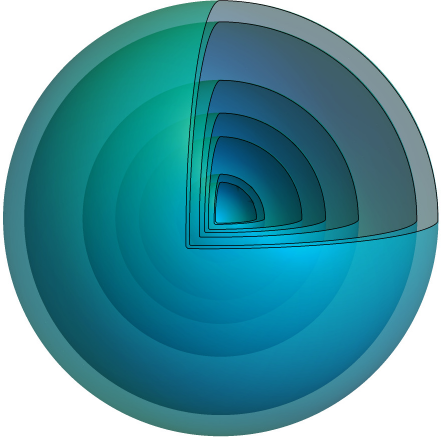} & \vspace{0.2cm}\includegraphics[height=1.5cm, keepaspectratio]{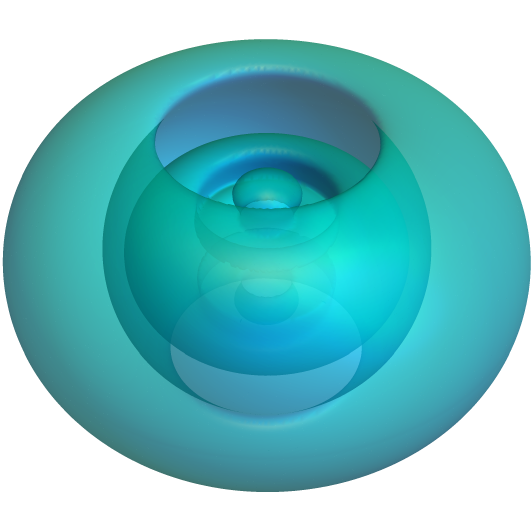} & \vspace{0.2cm}\includegraphics[height=1.5cm, keepaspectratio]{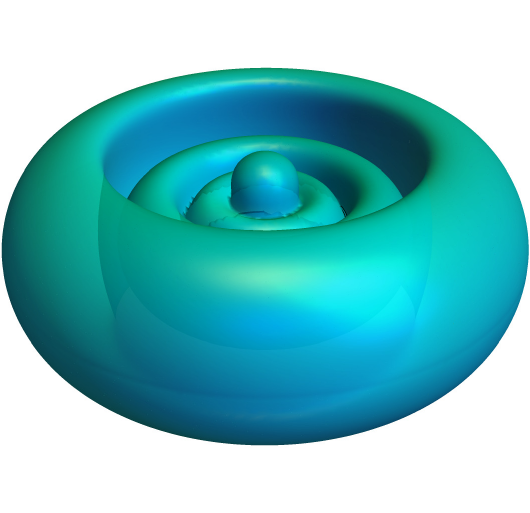} & \vspace{0.2cm}\includegraphics[height=1.2cm, keepaspectratio]{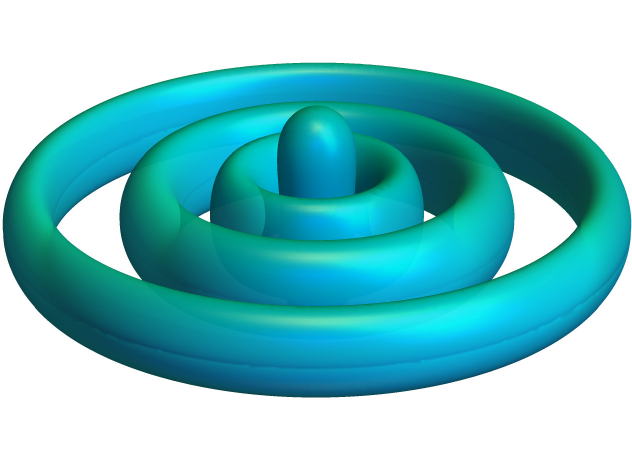} \\ \hline
  \end{tabular}
  \caption{Surfaces of constant energy density for scalar ABSs with $N_S=1,2,3$. The signs of $\epsilon$ indicate two directions bifurcation.  
  \label{density_scalar}}
\end{table}

\begin{table}[h!]
  \centering
  \begin{tabular}{|>{\centering\arraybackslash}m{0.4cm}|>{\centering\arraybackslash}m{1.7cm}|>{\centering\arraybackslash}m{1.7cm}|>{\centering\arraybackslash}m{1.7cm}|>{\centering\arraybackslash}m{1.7cm}|>{\centering\arraybackslash}m{1.7cm}|>{\centering\arraybackslash}m{1.7cm}|>{\centering\arraybackslash}m{1.7cm}|}
      \hline
      $N_S$\vspace{0.05cm} & 
      Chains\vspace{0.05cm} & 
      \multicolumn{2}{>{\centering\arraybackslash}m{3.4cm}|}{\vspace{0.3cm}\hspace*{0cm}$\overset{\text{\large $\epsilon > 0$}}{\xleftarrow{\hspace{2.8cm}}}$\hspace*{1cm}\vspace{0cm}} & 
      Spherical\vspace{0.05cm}&
      \multicolumn{2}{>{\centering\arraybackslash}m{3.4cm}|}{\vspace{0.3cm}\hspace*{0cm}$\overset{\text{\large $\epsilon < 0$}}{\xrightarrow{\hspace{2.8cm}}}$\hspace*{1cm}\vspace{0cm}} & 
      Rings\vspace{0.05cm} \\ \hline
      1 & 
      \vspace{0.2cm}\includegraphics[height=1.5cm, keepaspectratio]{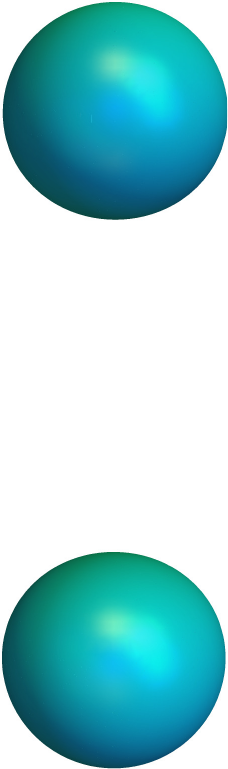} & 
      \vspace{0.2cm}\includegraphics[height=1.5cm, keepaspectratio]{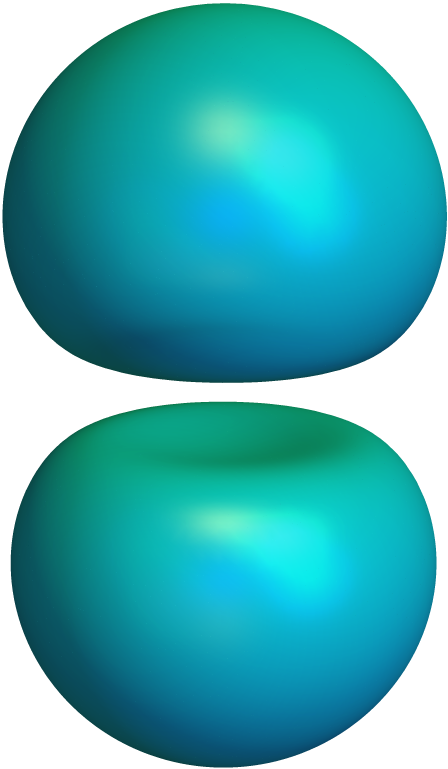} & 
      \vspace{0.2cm}\includegraphics[height=1.5cm, keepaspectratio]{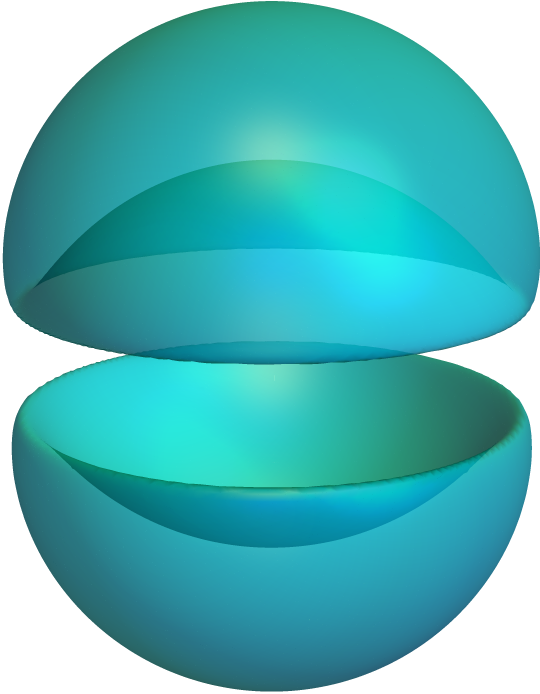} & 
      \vspace{0.2cm}\includegraphics[height=1.5cm, keepaspectratio]{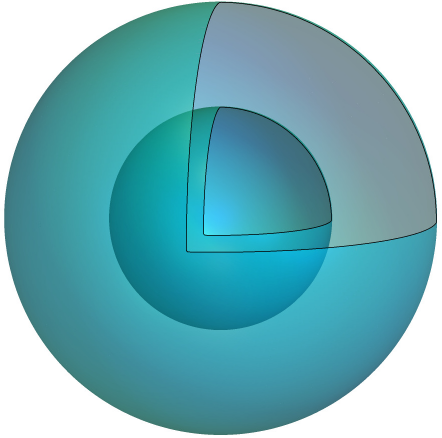} & 
      \vspace{0.2cm}\includegraphics[height=1.3cm, keepaspectratio]{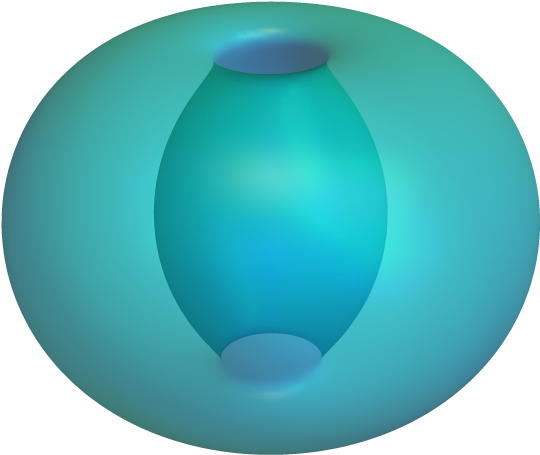} & 
      \vspace{0.2cm}\includegraphics[height=1.2cm, keepaspectratio]{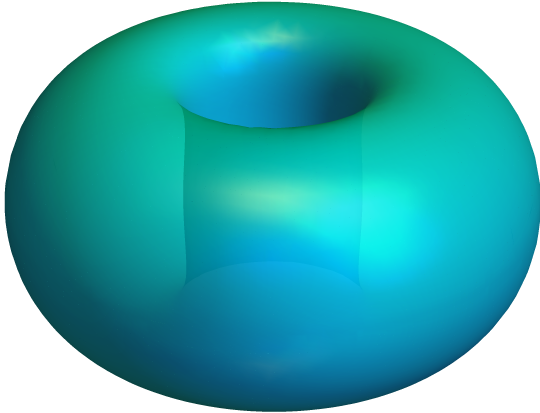} & 
      \vspace{0.2cm}\includegraphics[height=1cm, keepaspectratio]{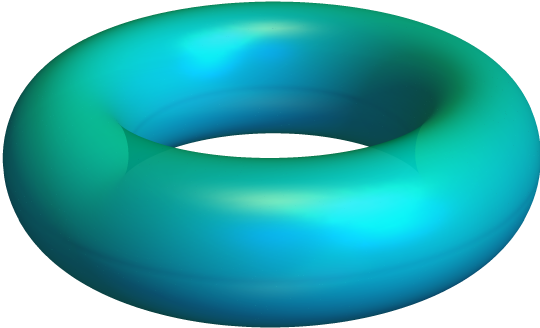} \\ \hline
  \end{tabular}
  \caption{Surfaces of constant energy density for Proca ABSs with $N_S=1$. The signs of $\epsilon$ are opposite to those of the scalar case in Table~\ref{density_scalar}.}
  \label{density_proca}
\end{table}
These bifurcations can also be demonstrated by the spatial distributions of the energy densities of the ABSs. As show in Table~\ref{density_scalar} and ~\ref{density_proca}, at bifurcation points, the surfaces of constant energy density for ABSs are spherically symmetric. By substituting the numerically solved perturbation functions and background solutions into (\ref{ansatz_pert}), and comparing them with the solutions of ABSs near the bifurcation points, we found that the sign of $\epsilon$ significantly affects the matter distribution in the system. For the scalar case, $\epsilon < 0$ reduces the energy density near the equatorial plane and increases it near the $z$-axis, causing matter to accumulate along the $z$-axis. The presence of nodes in the field functions results in noticeable fluctuations in the energy density along the $z$-direction, leading to a chain-like distribution of matter. In contrast, $\epsilon > 0$ leads to the accumulation of matter near the equatorial plane, forming a ring-like structure. Additionally, in the Proca case, the relationship between energy density and field functions is more complex, resulting in the effects of $\epsilon$ on Proca SBSs being opposite to those in the scalar field: $\epsilon > 0$ leads to a chain-like distribution, while $\epsilon < 0$ results in a ring-like distribution.

Our focus also extends to the quadrupole moment of the BSs.
In Newtonian gravity, the multipole moments of a system can be obtained through a multipole expansion of the gravitational field. In the context of relativistic gravity, the situation is more complex. Geroch \cite{Geroch:1970cd} and Hansen \cite{Hansen:1974zz} extended the concept of multipole moments from Newtonian gravity to general relativity, defining multipole moments in curved spacetime. We are particularly interested in the quadrupole moment, which can be measured through the gravitational waves emitted by compact bodies, and it reflects the symmetries of spacetime.

\begin{figure}[h!]
  \centering
  \includegraphics[width=0.49\textwidth]{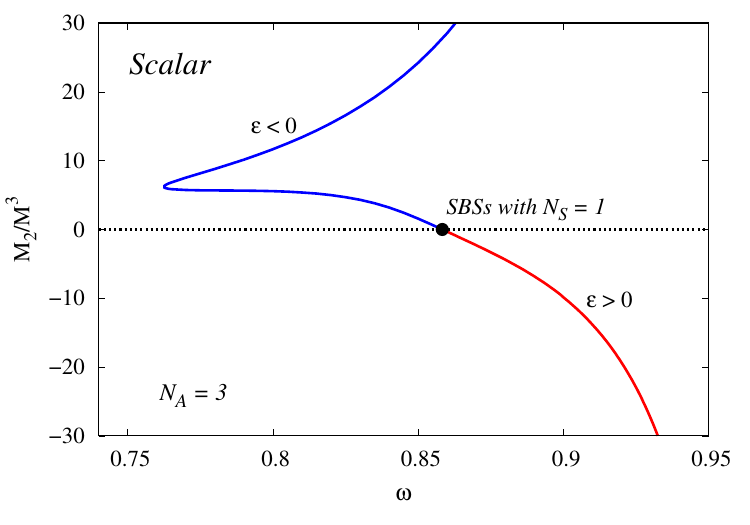}
  \includegraphics[width=0.49\textwidth]{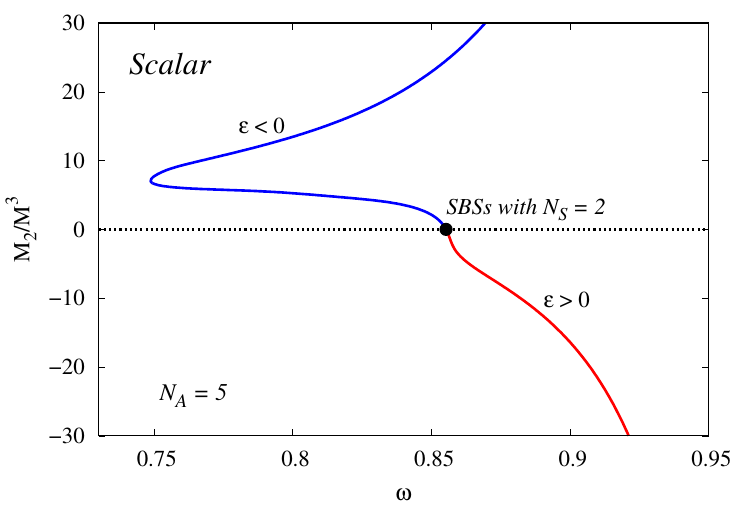}
  \includegraphics[width=0.49\textwidth]{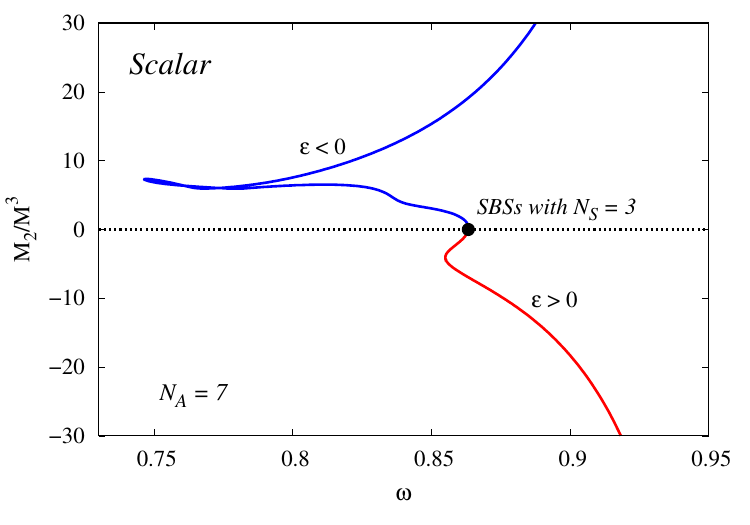}
  \includegraphics[width=0.49\textwidth]{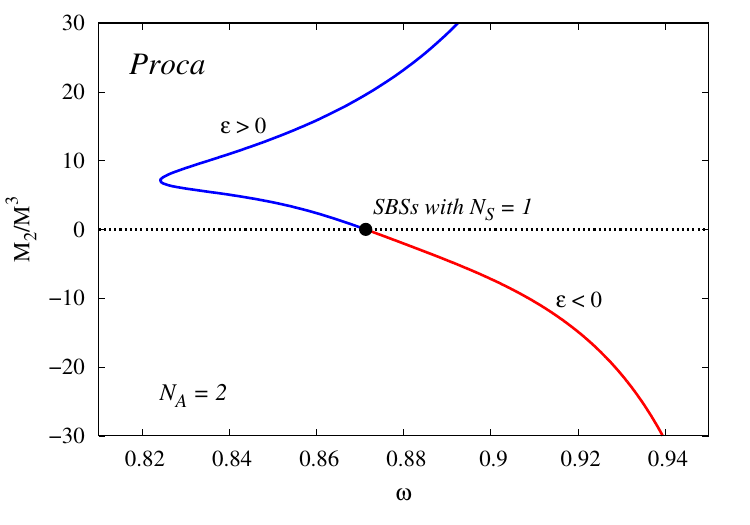}
  \caption{The quadrupole moments of the ABSs as functions of the frequency $\omega$.}
  \label{m2}
\end{figure}

The quadrupole moments of ABSs can be calculated using Eqs.~(\ref{quadrupole1})-(\ref{quadrupole4}). These physical quantities quantify the degrees of deviation from spherical symmetry and also indirectly reflect the connection between ABSs and SBSs. 
As shown in Fig.~\ref{m2}, the quadrupole moments of ABSs change with the frequency $\omega$, with the blue and red curves representing chain-like and ring-like ABSs, respectively. At the bifurcation points, indicated by the black dots, the models are spherically symmetric, thus $M_2 = 0$. We observe that perturbations with $\epsilon<0$ ($\epsilon>0$) induce chain-like scalar (Proca) ABSs branches with positive quadrupole moments, while perturbations with $\epsilon>0$ ($\epsilon<0$) induce ring-like scalar (Proca) ABSs branches with negative quadrupole moments. This means that a BS's quadrupole moment being greater than zero indicates a matter distribution closer to the z-axis; a quadrupole moment less than zero suggests that matter is accumulating near the equatorial plane.
%%%%%%%%%%%%%%%%%%%%%%%%%%%%%%%%%%%%%%%%%%%
\subsection{The results for $\ell = 4$ }
%%%%%%%%%%%%%%%%%%%%%%%%%%%%%%%%%%%%%%%%%%%
\begin{figure}[h!]
  \centering
  \includegraphics[width=0.6\textwidth]{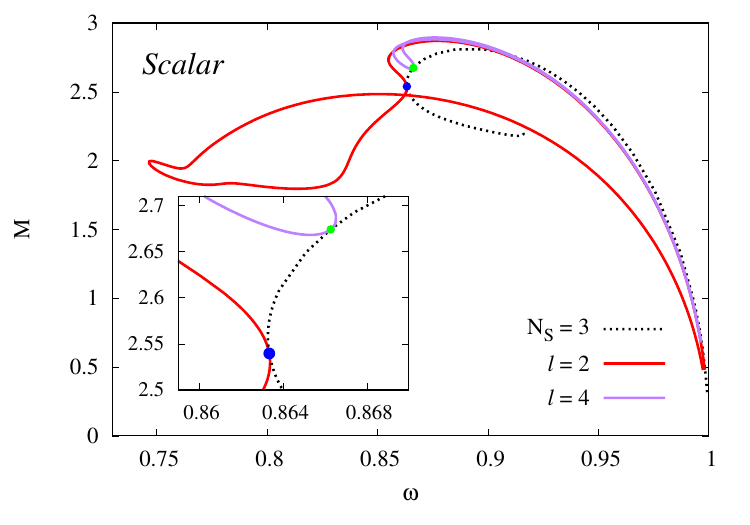}
  \caption{The ADM masses of the SBSs and ABSs associated with bifurcations under $\ell = 2$ and $\ell = 4$ perturbations}
  \label{m_l=4}
\end{figure}
Since the perturbation for $\ell=2$ is not special, zero-modes for $\ell \ne 2$ might exist. We found the bifurcations for $\ell \ne 2$ occur when $N_S$ is sufficiently large. Due to numerical challenges, we have only found one $\ell \ne 2$ bifurcation, which occurs in scalar SBSs with $N_S = 3$ under $\ell=4$ perturbation. The ADM mass of SBSs with $N_S=3$ and ABSs that have $\ell = 2$ and $\ell = 4$ zero-modes are shown in Fig.~\ref{m_l=4}. The black dashed line represents the SBSs with $N_S = 3$. The red and pueple curves represent the ABSs that have $\ell=2$ and $\ell=4$ bifurcations with the SBSs, respectively. The blue and green points are the background solutions where $\ell=2$ and $\ell= 4$ bifurcations occur.
\begin{figure}[h!]
  \centering
  \includegraphics[width=0.49\textwidth]{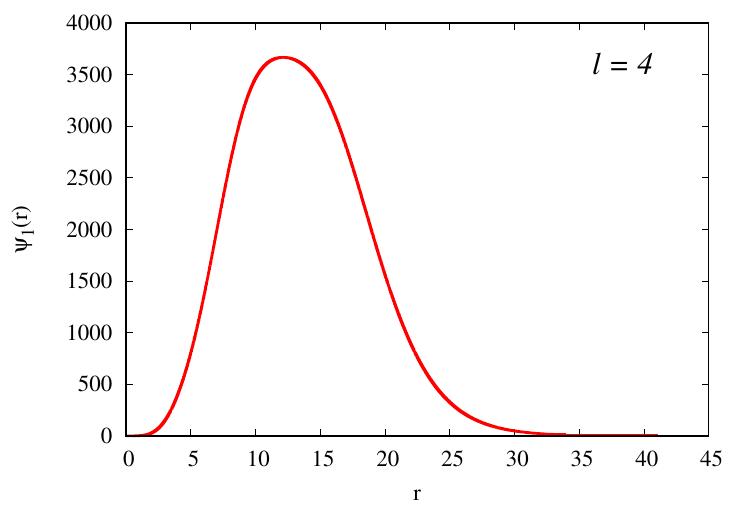}
  \includegraphics[width=0.49\textwidth]{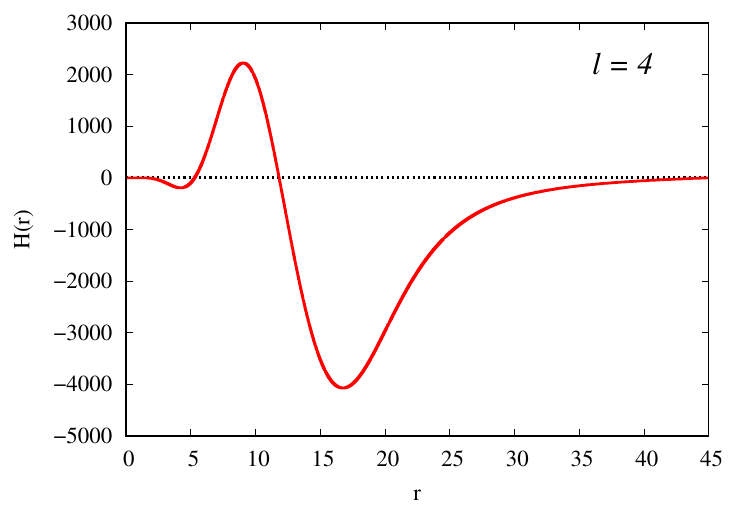}
  \caption{The profiles of perturbation functions $\psi_1(r)$ and $H(r)$ for $\ell=4$.}
  \label{h_psi1_l=4}
\end{figure}

\begin{table}[h!]
  \centering
  \begin{tabular}{|>{\centering\arraybackslash}m{0.4cm}|>{\centering\arraybackslash}m{1.7cm}|>{\centering\arraybackslash}m{1.7cm}|>{\centering\arraybackslash}m{1.7cm}|>{\centering\arraybackslash}m{1.7cm}|>{\centering\arraybackslash}m{1.7cm}|>{\centering\arraybackslash}m{1.7cm}|>{\centering\arraybackslash}m{1.7cm}|}
      \hline
      $N_S$\vspace{0.05cm} & 
      Mixed\vspace{0.05cm} & 
      \multicolumn{2}{>{\centering\arraybackslash}m{3.4cm}|}{\vspace{0.3cm}\hspace*{0cm}$\overset{\text{\large $\epsilon < 0$}}{\xleftarrow{\hspace{2.8cm}}}$\hspace*{1cm}\vspace{0cm}} & 
      Spherical\vspace{0.05cm}&
      \multicolumn{2}{>{\centering\arraybackslash}m{3.4cm}|}{\vspace{0.3cm}\hspace*{0cm}$\overset{\text{\large $\epsilon > 0$}}{\xrightarrow{\hspace{2.8cm}}}$\hspace*{1cm}\vspace{0cm}} & 
      Mixed\vspace{0.05cm} \\ \hline
      1 & 
      \vspace{0.2cm}\includegraphics[height=1cm, keepaspectratio]{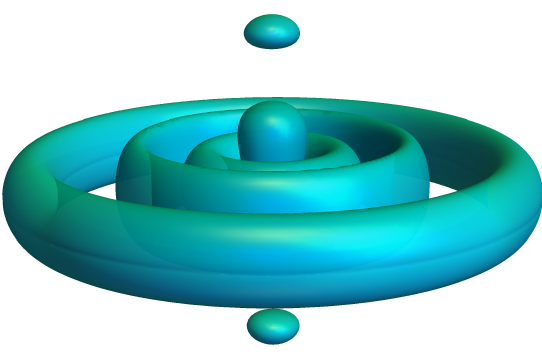} & 
      \vspace{0.2cm}\includegraphics[height=1.2cm, keepaspectratio]{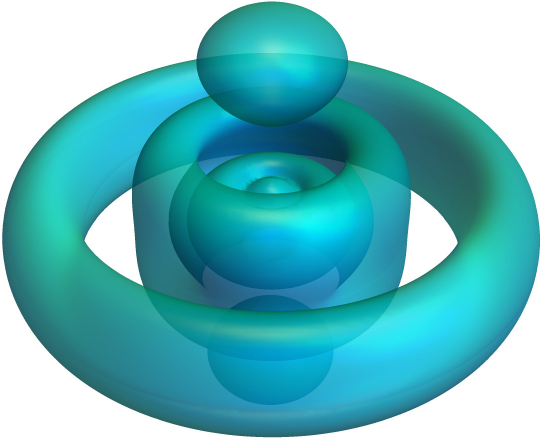} & 
      \vspace{0.2cm}\includegraphics[height=1.5cm, keepaspectratio]{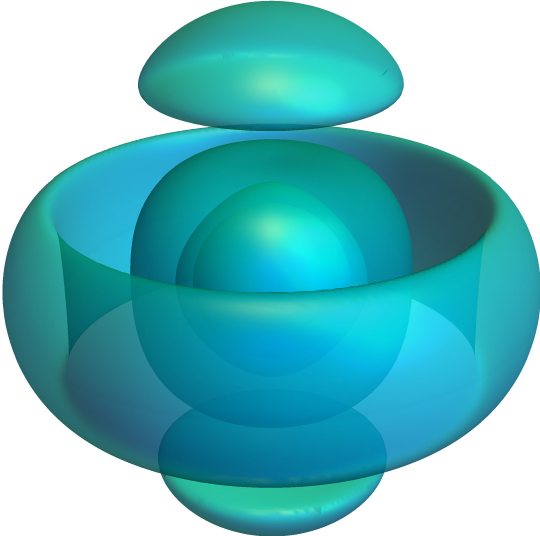} & 
      \vspace{0.2cm}\includegraphics[height=1.5cm, keepaspectratio]{plots/w3-eps-converted-to.pdf} & 
      \vspace{0.2cm}\includegraphics[height=1.3cm, keepaspectratio]{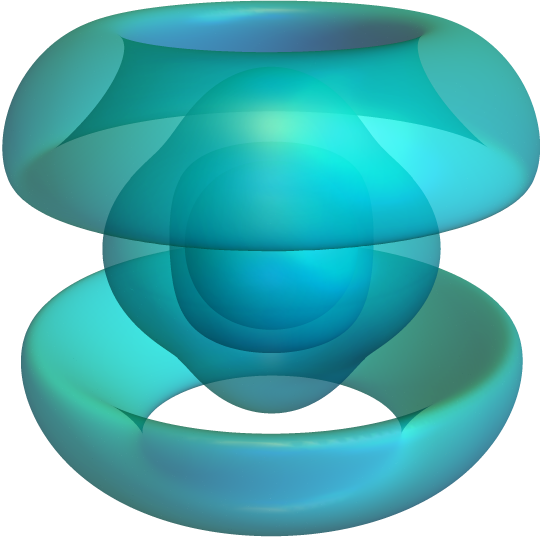} & 
      \vspace{0.2cm}\includegraphics[height=1.2cm, keepaspectratio]{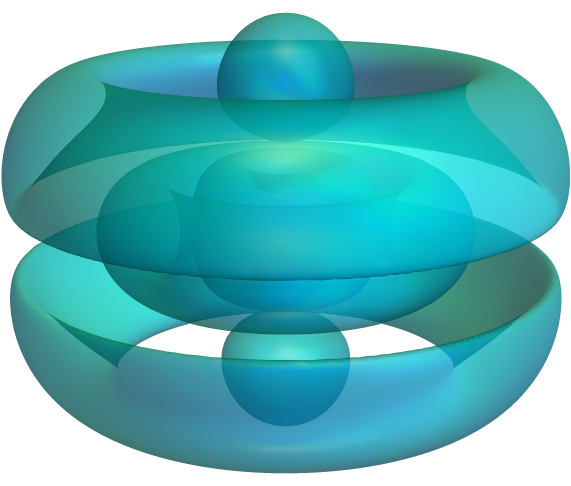} & 
      \vspace{0.2cm}\includegraphics[height=1cm, keepaspectratio]{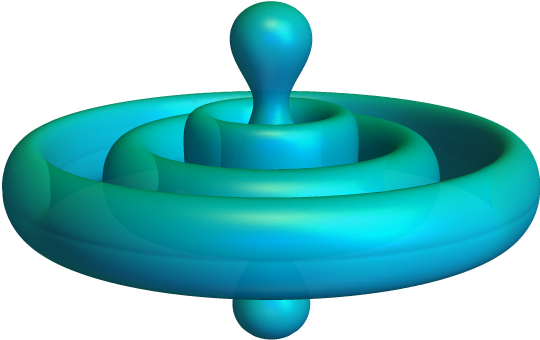} \\ \hline
  \end{tabular}
  \caption{Surfaces of constant energy density for scalar ABSs, which undergo $\ell = 4$ bifurcation with the SBSs.}
  \label{density_l=4}
\end{table}

\begin{figure}[h!]
  \centering
  \includegraphics[width=0.49\textwidth]{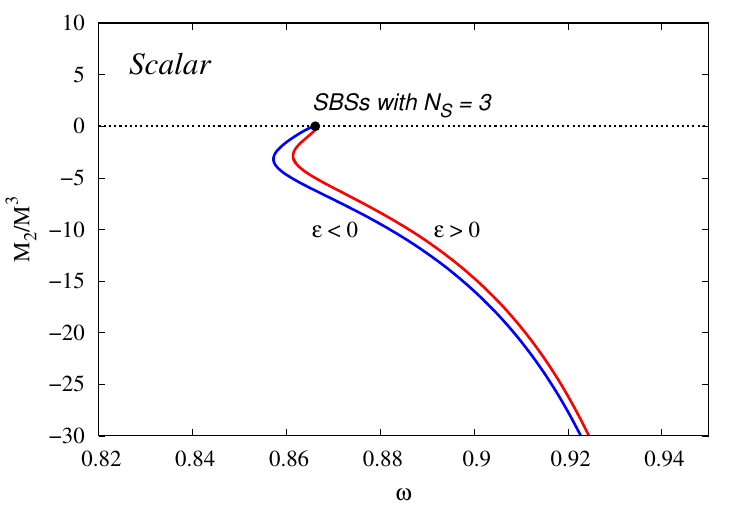}
  \caption{The quadrupole moments of the ABSs as functions of the frequency $\omega$ in $\ell=4$ case.}
  \label{m2_w_l=4}
\end{figure}

The profiles of the perturbation functions are shown in Fig.~\ref{h_psi1_l=4}. Compared to the $\ell = 2$ perturbations, the magnitudes of $\ell = 4$ perturbation functions are significantly higher, and the number of nodes in the perturbation functions also differs from those in $\ell = 2$ case. This indicates that the profiles of perturbation functions for different $\ell$ are difficult to predict, which poses a challenge in our search for zero-modes of $\ell=4$ perturbations.

Changes in matter distribution near the $\ell = 4$ bifurcation points are more complex than those near the $\ell = 2$ bifurcation points. Table~\ref{density_l=4} displays the two bifurcation scenarios for SBSs with $N_S=3$ when $\epsilon$ is positive and negative. Unlike $\ell = 2$ perturbation, the $\ell = 4$ perturbation causes the matter in the SBSs to accumulate along the $z$-axis and the equatorial plane. Although the solutions on either side of the bifurcation point (columns 3, 4, and 6, 7 in Table~\ref{density_l=4}) show significantly different matter distributions, the solutions closer to the Newtonian limit (columns 2 and 8) both exhibit a mix of chain-like and ring-like structures, resembling a gyroscope.

This type of gyroscope-like ABSs, compared to the ABSs mentioned above, also exhibits distinct characteristics in the quadrupole moment. As shown in Fig.~\ref{m2_w_l=4}, the quadrupole moment of the ABSs is always less than zero. This also reflects, to some extent, the characteristics of the matter distribution of the ABSs. Although the matter distribution appears as a mixture of chain-like and ring-like structures in Table~\ref{density_l=4}, the negative quadrupole moment indicates that there is a more significant accumulation of matter near the equatorial plane, while matter near the z-axis is relatively sparse, resulting in a gyroscope-like matter distribution.

%%%%%%%%%%%%%%%%%%%%%%%%%%%%%%%%%%
\section{Conclusions}\label{conclusions}
%%%%%%%%%%%%%%%%%%%%%%%%%%%%%%%%
In this work, we have studied the bifurcation phenomena within the BSs and explained these bifurcations using axisymmetric perturbation theory. We discovered that exhibit zero-modes under axisymmetric perturbations at $\ell=2$ and 
$\ell=4$, which lead to a series of ABSs with unique matter distribution forms, including the chain-like, ring-like, and the gyroscope-like ABSs we constructed for the first time. Additionally, we extended the chains of BSs to configurations with seven components.

We observed that the SBSs with $N_S>0$  bifurcate in two opposite directions under axisymmetric perturbations, leading to different matter distribution patterns. The direction of bifurcation depends on the sign of the small parameter $\epsilon$, similar to the bifurcation of charged drops going beyond the Rayleigh limit as shown in Fig.~2 of \cite{Basaran:1989cd}. Furthermore, the ABSs resulting from bifurcations of SBSs share common characteristics. The relationship between their ADM mass and frequency forms loops, with each branch close to the Newtonian limit corresponding to either chain-like or ring-like ABSs. As the frequency changes, matter initially further from the center gradually converges towards the central region, forming multiple spherical shells. Ultimately, the ABSs transform into the SBSs at the bifurcation points.

A direct extension of this work involves considering the introduction of self-interactions. Studies in \cite{Sanchis-Gual:2021phr,Brito:2023fwr} on the dynamical stability of SBSs with quartic self-interactions suggest that a sufficiently strong self-interaction can stabilize the SBSs with $N_S>0$. We speculate that the self-interactions coupling constant $\lambda$ has a threshold $\lambda_0$ related to the zero-modes of axisymmetric perturbations, and a self-interaction with $\lambda > \lambda_0$ could potentially eliminate the bifurcations in the SBSs.

An interesting question in the context of this work is
 whether other configurations, in particular those with discrete symmetries \cite{Herdeiro:2020kvf},
may emerge as zero modes of other 
BSs.
%possessing an higher degree of symmetry.
Here it is 
  noteworthy to remark that the boundary conditions significantly limit the possibility of bifurcations between different BSs, hence SBSs cannot bifurcate into any solution that does not match their boundary conditions under perturbations, such as those in \cite{Herdeiro:2020kvf} except the {\it axially symmetric odd-chains}.
  Moreover, we have explored potential bifurcations among ABSs by constructing other static ABSs  apart from  those mentioned in this work or those in Ref.  \cite{Herdeiro:2020kvf}
  (but still sharing the same boundary conditions), and analyzed the results. However, so far we did not find any indication of bifurcations within these ABSs. Nevertheless, the possibility of new bifurcations cannot be ruled out. 
 Here one should remark that,
 for  the nomenclature of Ref. \cite{Herdeiro:2020kvf}, the
 scalar ABSs in this work correspond to 
  {\it hybrid} multipolar BSs ,
 and can be interpreted as superpositions of different
 stationary states.
Therefore,
   a promising direction for future study would be to consider new families of
   hybrid multipolar BSs
   by combining different  types of ABSs, thereby possibly uncovering new  bifurcations.

We can also introduce a $U(1)$ gauged field to examine its impact on the stability of charged SBSs under axisymmetric perturbations, explore potential zero-modes, and attempt to construct related charged ABSs. Taking into account the spherical charged BHs with $Q$-hair detailed in \cite{Herdeiro:2020xmb,Hong:2020miv}, we are poised to further investigate the zero-modes of such models under axisymmetric perturbations, aiming to use these insights to attempt constructing charged multi-BH solutions with $Q$-hair.

%%%%%%%%%%%%%%%%%%%%%%%%%%%%%%%%%%%%%%%%%%%%%%%%%%%%%%%%%%%%%%%%%%
\section*{Acknowlegements}
%%%%%%%%%%%%%%%%%%%%%%%%%%%%%%%%%%%%%%%%%%%%%%%%%%%%%%%%%%%%%%%%%%
C. L. are supported by China Scholarship Council (CSC) fellowship. This work is supported by the Center for Research and Development in Mathematics and Applications (CIDMA) through the Portuguese Foundation for Science and Technology (FCT -- Fundac\~ao para a Ci\^encia e a Tecnologia) under the Multi-Annual Financing Program for R\&D Units, PTDC/FIS-AST/3041/2020 (\url{http://doi.org/10.54499/PTDC/FIS-AST/3041/2020}),  2022.04560.PTDC (\url{https://doi.org/10.54499/2022.04560.PTDC}) and 2024.05617.CERN. This work has further been supported by the European Horizon Europe staff exchange (SE) programme HORIZON-MSCA-2021-SE-01 Grant No.\ NewFunFiCO-101086251. 
%%%%%%%%%%%%%%%%%%%%%%%%%%%%%%%%%%%%%%%%%%%%%%%%%%%%%%%%%%%%%%%%%%
\begin{small}

\end{small}

 \end{document}